\documentclass[twocolumn, prx, reprint, amsfonts, amsmath, amssymb, superscriptaddress, longbibliography, floatfix]{revtex4-2}
\usepackage[T1]{fontenc}
\usepackage{graphicx}
\usepackage[dvipsnames]{xcolor}
\usepackage[colorlinks,linkcolor=Blue,citecolor=Blue]{hyperref}
\usepackage{braket}

\begin{document}

\title{The Fermi-Dirac staircase occupation of Floquet bands and current rectification inside the optical gap of metals: a rigorous perspective}

\author{Oles Matsyshyn}
\affiliation{Max-Planck-Institut f{\"u}r Physik komplexer Systeme, N{\"o}thnitzer Stra{\ss}e 38, 01187 Dresden, Germany}
\affiliation{Division of Physics and Applied Physics, School of Physical and Mathematical Sciences, Nanyang Technological University, Singapore 637371, Republic of Singapore}

\author{Justin C. W. Song}
\affiliation{Division of Physics and Applied Physics, School of Physical and Mathematical Sciences, Nanyang Technological University, Singapore 637371, Republic of Singapore}

\author{Inti Sodemann Villadiego}
\email[sodemann@itp.uni-leipzig.de]{}
\affiliation{Institut f{\"u}r Theoretische Physik, Universit{\"a}t Leipzig, Br{\"u}derstra{\ss}e 16, 04103, Leipzig, Germany}
\affiliation{Max-Planck-Institut f{\"u}r Physik komplexer Systeme, N{\"o}thnitzer Stra{\ss}e 38, 01187 Dresden, Germany}

\author{Li-kun Shi}
\email[shi@itp.uni-leipzig.de]{}
\affiliation{Max-Planck-Institut f{\"u}r Physik komplexer Systeme, N{\"o}thnitzer Stra{\ss}e 38, 01187 Dresden, Germany}
\affiliation{Institut f{\"u}r Theoretische Physik, Universit{\"a}t Leipzig, Br{\"u}derstra{\ss}e 16, 04103, Leipzig, Germany}

\date{\today}

\begin{abstract}
We consider a model of a Bloch band  subjected to an oscillating electric field and coupled to a featureless fermionic heat bath, which can be solved exactly.
We demonstrate rigorously that in the limit of vanishing coupling to this bath (so that it acts as an ideal thermodynamic bath) the occupation of the Floquet band is not a simple Fermi-Dirac distribution function of the Floquet energy, but instead it becomes a ``staircase'' version of this distribution.
We show that this distribution generically leads to a finite rectified electric current within the optical gap of a metal even in the limit of vanishing carrier relaxation rates, providing a rigorous demonstration that such rectification is generically possible and clarifying previous statements in the optoelectronics literature.
We show that this current remains non-zero even up to the leading perturbative second order in the amplitude of electric fields, and that it approaches the standard perturbative expression of the Jerk current obtained from a simpler Boltzmann description within a relaxation time approximation when the frequencies are small compared to the bandwidth.
\end{abstract}

\maketitle

\section*{Introduction}

Quantum many body systems that are periodically driven in time have garnered attention over the last decades as rich platforms to realize novel collective phenomena and non-equilibrium states beyond those realized in equilibrium~\cite{oka2009photovoltaic,kitagawa2010topological,kitagawa2011transport,lindner2011floquet,jiang2011majorana,rudner2013anomalous,ponte2015many,lazarides2015fate,carpentier2015topological,abanin2016theory,khemani2016phase,else2016floquet,yao2017discrete,choi2017observation,zhang2017observation,titum2016anomalous,potter2016classification,else2016classification,von2016phase,roy2017floquet,po2016chiral,po2017radical,nathan2019anomalous,aleiner2021bethe,morvan2022formation}.
A phenomenon that can arise in such periodically driven systems and which is forbidden in equilibrium, is the existence of an average net rectified particle flow or ratchet effect.
In particular for the case of electrons in crystals driven by oscillating electric fields, these effects have enjoyed recent renewed attention due to their interesting interplay with the dispersions and Berry phases of band structures and their potential for novel opto-electronic devices~\cite{young2012first,matsyshyn2021berry,brehm2014first,rangel2017large,cook2017design,morimoto2018current,kumar2021room}.
Despite their natural connection, only a handful of studies have studied current rectification effects in Bloch bands through the lens of Floquet theory~\cite{morimoto2016topological,morimoto2016semiclassical,matsyshyn2021rabi}.
This is in part due to the difficulty that even for non-interacting systems, there is no simple general formula dictating the occupation of Floquet bands analogous to the Fermi-Dirac distribution that dictates the occupation of bands in equilibrium~\cite{bilitewski2015scattering,shirai2015condition,seetharam2015controlled,genske2015floquet,shirai2016effective,esin2018quantized}.
\begin{figure}[t]
\includegraphics[width=0.48\textwidth]{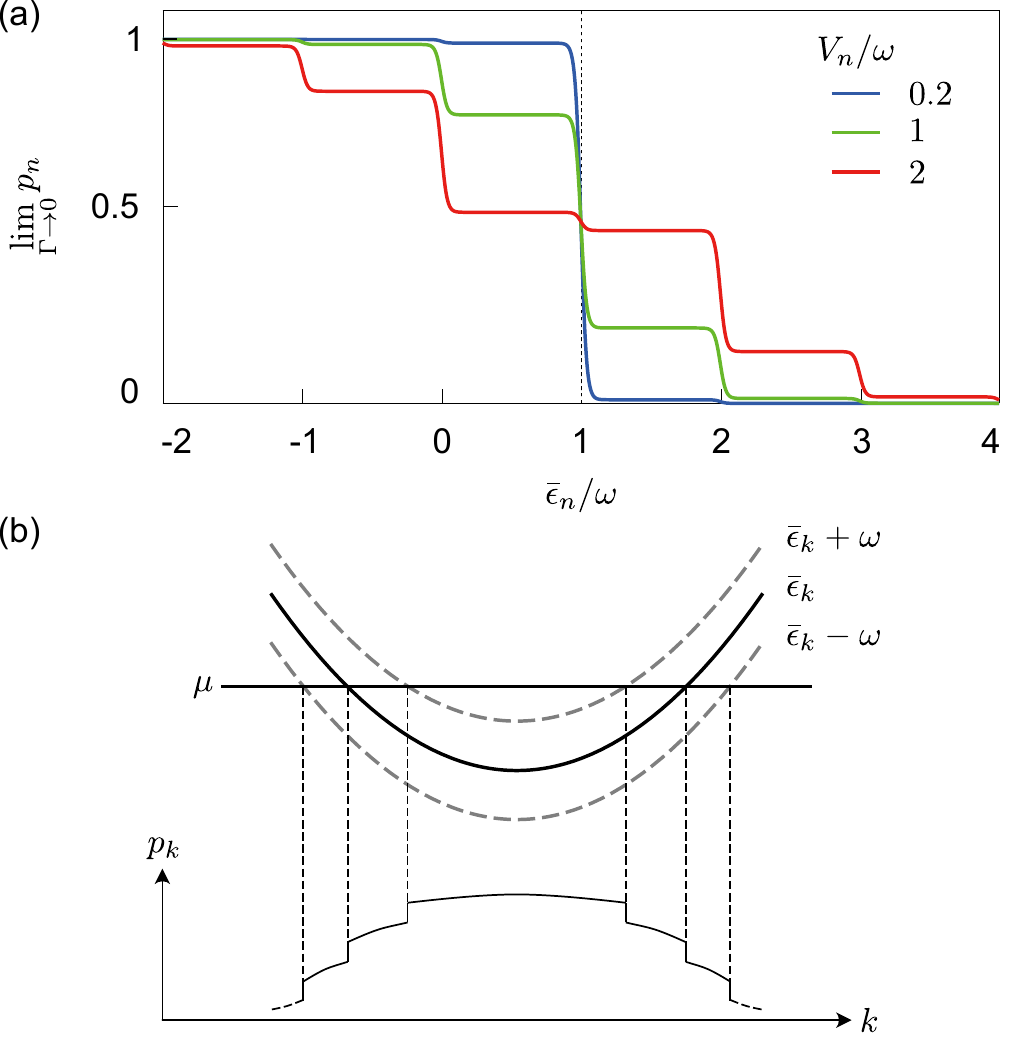}
\caption{
(a) Time independent $\lim_{\Gamma \to 0} p_n$ as a function of $\bar{\epsilon}_n / \omega$ calculated using Eq.~(\ref{p_n_average}) showing a ladder-like occupation. Parameters used: $\beta/\omega = 50$, $\mu/\omega = 1$.
(b) Schematic of the ladder-like occupation for a Bloch band.}
\label{fig-ladder}
\end{figure}

One of the central goals of our study is to provide simple analytical formulae for the occupation of a single Floquet band coupled to a ``featureless fermionic bath'' (which is a commonly used model of bath that for example was employed in Refs.~\cite{morimoto2016topological,matsyshyn2021rabi,park2022thermal}).
This featureless bath is a physical system that has a finite coupling to the fermionic system of interest and is characterized by a single relaxation scale $\Gamma$.
The dynamics of the system coupled to this bath can be described in an exact manner thanks to the fact that we will take the combined system plus bath as a non-interacting fermionic system.
As we will see, this featureless fermionic bath behaves as an ideal thermodynamic bath in the limit in which its coupling to system is vanishingly small ($\Gamma \to 0$), and in particular in this limit it relaxes the system towards the equilibrium Fermi-Dirac occupation of the bands in the absence of an external periodic drive.
As we will show, however, when the system is periodically driven in time, this bath leads to a self-consistent occupation the Floquet bands that is sharply different from that of the equilibrium Fermi-Dirac distribution, but which we can determine analytically with no approximations.
This occupation is instead a staircase version of the Fermi-Dirac distribution with several jumps that occur at copies of the chemical potential shifted by all the harmonics of the driving frequency [see Fig.~\ref{fig-ladder}(a)].
In the limit of an ideal bath ($\Gamma \to 0$), we will show that this distribution coincides with the distribution that has been previously obtained within the Boltzmann approach to Floquet systems (see in particular Eq.~(12) of Ref.~\cite{seetharam2015controlled}).

Another central purpose of our study is to exploit the Floquet formalism to further elaborate on our recent finding~\cite{shi2022berry} that it is indeed possible for time dependent oscillating electric fields with a frequency that lies within the optical gap of a metal, to induce a net rectified DC electric current.
We will see that this is true even when the electric field has a single monochromatic frequency $\omega$ that is much larger than the relaxation rates and this current remains finite in the limit when these rates vanish ($\Gamma \to 0$) (and therefore does not rely on the frequency difference effect~\cite{de2020difference} or in the Raman scattering effect~\cite{onishi2022effects,golub2022raman}).
We will demonstrate that this is possible by choosing a simple model containing a single Bloch band with no Berry curvature, which in a simpler relaxation time Boltzmann description would give rise to the so-called ``Jerk'' effect described in Refs.~\cite{matsyshyn2019nonlinear,shi2022berry}.
Our aim is to use this simple model because it will allow us to carry out calculations of its response coupled to the fermionic bath in a clear and exact analytical manner.

We are motivated to do this rigorously in order to 
clarify a 
series of misconceptions that originated from the work of
Belinicher, Ivchenko and Pikus~\cite{belinicher1986transient} and that have propagated into some of the subsequent literature
~\cite{ivchenko1988magneto,onishi2022effects,golub2022raman,pershoguba2022direct,shi2022berry}.
In Appendix~\ref{appendixE} we comment in more detail about some of these previous works and point out more specifically some of their imprecisions.

One of the central messages of our study is that it is indeed possible to have a net rectified current in response to a monochromatic oscillating electric field whose frequency lies within the optical gap of a system, in the limit of vanishing carrier relaxation rates.
We will demonstrate this within a self-consistent picture of the occupation of Floquet Bloch band in the steady state of the system. More specifically we will show that in the limit of an ideal bath ($\Gamma \to 0$) the average rectified current in the non-equilibrium steady state of the system is given by:
\begin{align}
& \bar{\bf j} 
= \int_{\bf k}
p_{\bf k} 
\nabla_{\bf k} \bar{\epsilon}_{\bf k} ,
\end{align}
where $\bar{\epsilon}_{\bf k}$ is the Floquet energy of the band, $p_{\bf k} $ is the occupation of the Floquet band, $\bar{\bf j} $ is the current averaged over one period, and the integral is over the crystal momentum in the Brillouin zone with the usual normalization of $1/(2\pi)^{d}$.



The crucial difference between the above expression and that for the average current in an equilibrium system, is that the occupation function $p_{\bf k}$ is not simply the Fermi-Dirac distribution associated with $\bar{\epsilon}_{\bf k}$, but instead it is precisely the stair-case occupation function depicted in Fig.~\ref{fig-ladder}.
Crucially, we will show that generically this stair-case occupation is a function that depends on all the information of the time dependence of the Hamiltonian, and cannot be expressed as a function of only $\bar{\epsilon}_{\bf k}$.
We will then show that as a consequence of this, the rectified current is in fact generically non-zero in the optical gap of a metal that breaks inversion and time-reversal symmetries.
This result remains true even to second order in the amplitude of the oscillating electric field, which is the leading order at which rectification currents appear, and therefore implies a non-vanishing rectification conductivity within the optical gap of metals, in agreement with our previous results~\cite{shi2022berry}.
For other previous discussions of the possibility of in-gap rectification see also Refs.~\cite{sodemann2015quantum,kaplan2020nonvanishing,gao2021intrinsic,watanabe2021chiral,golub2022raman,golub2020semiclassical}.

Our paper is organized as follows.
In Section~\ref{section-1}, we setup the approach to open quantum systems, obtain exact occupation functions for diagonal and time-periodic Hamiltonians coupled to a featureless fermionic bath.
In Section~\ref{section-2}, based on the exact occupation functions, we calculate exact linear and rectification conductivities and show that there is a net rectified current in response to a monochromatic oscillating electric field whose frequency lies within the optical gap of a metal, in the limit of vanishing carrier relaxation rates.

\section{The open-system Schr{\"o}dinger Equation Approach to open quantum systems}
\label{section-1}

In descriptions of quantum open systems it is typically natural to view the combined Hilbert space of the ``system'' and the ``bath'' as a tensor {\it product} of their Hilbert spaces in isolation.
There are situations, however, where it is possible to alternatively cast this separation of system and bath as a direct {\it sum} of their individual Hilbert spaces.
As we will show, such separation into sums of Hilbert spaces is extremely powerful and convenient, because it allows to integrate out the dynamics of the ``bath'' in an exact manner and to obtain a simple non-Hermitian generalization of Schr{\"o}dinger's equation for the system which captures its coupling to the bath without any approximations.

One example of the class of models which admits such direct sum separation into system and bath are those of non-interacting particles.
To see this let us imagine that the system and the bath as a whole can be described by a non-interacting model.
For concreteness we can imagine this to be a tight biding model of particles hopping on a lattice. Because the problem is non-interacting, then the dynamics can be analyzed by computing the trajectories of single individual particles and then adding them up.
However, for a single particle the Hilbert space of the ``system'' and the ``bath'' can be naturally viewed as a direct sum.
For example, in the case of a tight-binding model, some sites can be viewed as belonging to the system and the remainder sites as belonging to the bath.

Let us then consider that the Hilbert space of the system and the bath can be decomposed into a direct sum, namely their Hamiltonian and states have block form as follows:
\begin{align}
H (t) = \begin{bmatrix}
H_S (t) & H_{SB} (t) \\
H_{BS} (t) & H_B (t)
\end{bmatrix} ,
\quad
\ket{\psi(t)} = \begin{bmatrix}
\ket{\psi_S (t)}  \\
\ket{\psi_B (t)} 
\end{bmatrix} ,
\label{Full-Hamiltonian}
\end{align}
where $H_{BS} (t) = H_{SB}^\dagger (t)$.
From Eq.~(\ref{Full-Hamiltonian}), the coupled Schr\"{o}dinger equations for system and bath states then read:
\begin{align}
i \partial_t \ket{\psi_S (t)} = H_S (t) \ket{\psi_S (t)} + H_{SB} (t) \ket{\psi_B (t)} ,
\label{SEforSystem}
\\
i \partial_t \ket{\psi_B (t)} = H_{BS} (t) \ket{\psi_S (t)} + H_{B} (t) \ket{\psi_B (t)} ,
\label{SEforBath}
\end{align}
where we set $\hbar = 1$ throughout the paper.
By integrating Eq.~(\ref{SEforBath}) over time and inserting it into Eq.~(\ref{SEforSystem}) allows to formally eliminate the bath state dynamics $\ket{\psi_B (t)} $ and to obtain the open-system Schr\"{o}dinger equation for $\ket{\psi_S (t)}$:
\begin{align}
i \partial_t & \ket{\psi_S (t)} = H_S (t) \ket{\psi_S (t)} + H_{SB} (t) U_B (t, t_0) \ket{\psi_B (t_0)} 
\nonumber\\
& \quad - i H_{SB} (t) \int_{t_0}^t {\text d} t' ~ U_B (t,t') H_{BS} (t') \ket{\psi_S (t')} ,
\label{OSE-0}
\end{align}
where $U_B (t,t') $ is the bath (intrinsic) evolution operator satisfying $i \partial_t U_B (t,t_0) = H_B (t) U_B (t,t_0) $.
This procedure is often carried within the Schwinger-Keldysh formalism by integrating out part of the action describing the degrees of freedom of the bath (see e.g. Refs.~\cite{morimoto2016topological,matsyshyn2021rabi,park2022thermal}).
But this is easier and more physically transparent in our first quantization notation and the final results would be identical.

\subsection{Featureless fermionic bath}
We will now specialize the above equation to a model of a ``featureless fermionic bath'', which we define as having the following characteristics:

\begin{figure}[b]
\includegraphics[width=0.48\textwidth]{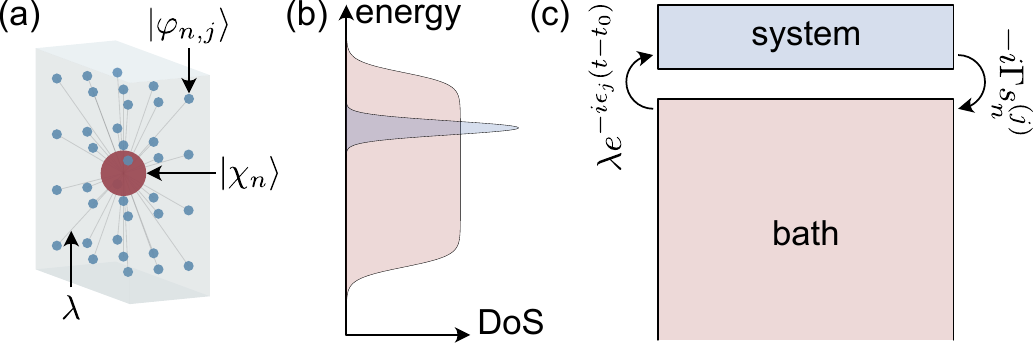}
\caption{(a) Schematic of the system-bath coupling $H_{SB}$.
(b) The bath's density of states is much wider than that of the system, and we simplify it to be flat in the energy range of interest.
(c) Schematic of the bath acting as a source as well as a sink for the system [see Eq.~(\ref{OSE-5})].}
\label{fig-system-bath}
\end{figure}
(i) In a featureless fermionic bath every state of the system is coupled to a collection of identical sites with the same energy spectrum and the same coupling $\lambda$ [see Fig.~\ref{fig-system-bath}(a)].
If the system states (basis) are denoted by $\ket{\chi_n}$ and the bath states (basis) by $\ket{\varphi_{n,j}}$, the bath and the system-bath coupling are
\begin{align}
& H_{B} = \sum_{n, j} \varepsilon_j \ket{\varphi_{n, j}}\bra{\varphi_{n, j}} ,
\label{HB}
\\
& H_{SB} = \lambda \, \sum_{n, j} \ket{\chi_{n}} \bra{\varphi_{n, j}} ,
\label{HSB}
\end{align} 
where $\varepsilon_j $ is the energy for the bath state $\ket{\varphi_{n,j}}$.
This model of the bath is identical to that employed in Refs.~\cite{shi2022berry,nagaosa2017concept,gerchikov1989theory,fregoso2013driven,kamenev2004many,johnsen1999quasienergy,jauho1994time,kohler2005driven,matsyshyn2021rabi}.

(ii) The featureless fermionic bath is prepared in an initial condition at $t_0$ with a Fermi-Dirac distribution that only has weight on the bath sites, described by
\begin{align}
& \rho_S (t_0) = 0 , \quad
\rho_B (t_0) = { \textstyle \sum_{n, j} }  f_0 (\varepsilon_j) \ket{\varphi_{n,j}}\bra{\varphi_{n,j}} ,
\label{initial-rho}
\\
& f_0 (\varepsilon_j) = \frac{1}{\exp[\beta_0 (\varepsilon_j - \mu_0 )]+1},
\end{align}
where $\mu_0$ is the chemical potential of and $\beta_0 = 1/ k_B T_0 $ denotes the temperature of the bath, respectively.
The assumption of the initial density matrix only having weight on the bath is useful but it is not strictly necessary.
This is because in the limit in which the bath spectrum becomes a dense continuum, the information of the initial condition for the component of density matrix on the system will decay over time and only the information of the initial condition for the density matrix on the bath will dictate the late time steady state [this will become more clear in Eq.~(\ref{OSE-2}) which is a subsequent version of Eq.~(\ref{OSE-0})].
Notice also that we have equated the evolution of the single particle density matrix with that of the many-body one-particle density matrix (equal time Greens function), which is possible thanks to the fact that the system is non-interacting.

With assumptions (i) and (ii), by evolving the initial condition in Eq.~(\ref{initial-rho}) under Eqs.~(\ref{SEforSystem}) and (\ref{SEforBath}), one finds that the one-body density matrix projected onto the system at time $t$ is given by
\begin{align}
\rho_S (t) = \sum_{n, j} f_0 (\varepsilon_j) \ket{\psi_n^{(j)} (t)} \bra{\psi_n^{(j)} (t)} ,
\label{System-Occupation-0}
\end{align}
where $\ket{\psi_n^{(j)} (t)}$ is the component within system Hilbert space that evolves out of the initial state $\ket{\psi_B (t_0)} = \ket{\varphi_{n,j}}$ in the bath at $t_0$. Eq.~(\ref{System-Occupation-0}) states that the density matrix for the system is the weighted sum of contributions from all bath states with their corresponding initial occupations.

Using Eqs.~(\ref{OSE-0}), (\ref{HB}), and (\ref{HSB}), we obtain the open-system Schr{\"o}dinger equation for $\ket{\psi_n^{(j)} (t)}$:
\begin{align}
i \partial_t \ket{\psi_n^{(j)} (t)} 
& = H_S (t) \ket{\psi_n^{(j)} (t)} + \lambda \exp[- i \varepsilon_j (t-t_0) ] \ket{\chi_{n}}
\nonumber\\
& \qquad - i \int_{t_0}^\infty {\text d} t' ~ \gamma (t - t') \ket{\psi_n^{(j)} (t')} .
\label{OSE-1}
\end{align}
Here, $\lambda \exp[- i \varepsilon_j (t-t_0) ] \ket{ \chi_{n} }$ is a source term for $\ket{ \psi_n^{(j)} (t) }$ arising from the bath, while the memory function in the second line is given by,
\begin{align}
\gamma (t) = \lambda^2 \Theta (t) \sum_{j} \exp (- i \varepsilon_j t) ,
\end{align}
which encodes the memory of decay for $\ket{ \psi_n^{(j)} (t) }$ due to the bath.
This memory function makes the Schr{\"o}dinger equation for open systems non-local in time, and in general it incorporates decay and renormalizations of the system energies due to their coupling to the bath [see Fig.~\ref{fig-system-bath} for a depiction].

(iii) To remove the finite memory delay, we impose one further property defining the featureless fermionic bath, namely that it has an infinitely broad and flat spectrum [see Fig.~\ref{fig-system-bath}(b)], i.e., the bath density of state is constant:
\begin{align}
\nu_B (\omega_b) = 2\pi \sum_{j} \delta (\omega_b - \varepsilon_j ) \equiv  \nu_0 .
\label{broad-flat-bath}
\end{align}
With this simplification, the finite delay or non-local memory of the past time $t'$ in Eq.~(\ref{OSE-1}) is lost, the memory function becomes:
\begin{align}
\gamma (t) 
& = \lambda^2 \Theta (t) \sum_{j} \int_{-\infty}^{+\infty} {\text d} \omega_b ~ \delta (\omega_b - \varepsilon_j ) \exp (- i \omega_b t)
\nonumber\\
& = \lambda^2 \nu_0 \Theta (t)  \int_{-\infty}^{+\infty} 
\frac{ {\text d} \omega_b}{ 2\pi }  \exp (- i \omega_b t)
= \delta (t) \, \Gamma ,
\end{align}
where we used Eq.~(\ref{broad-flat-bath}) to obtain the second equation and defined
\begin{align}
\Gamma \equiv \frac{\lambda^2 \nu_0}{ 2 } .
\label{Gamma-definition}
\end{align}
With the above simplification of infinitely broad spectrum for the bath, the open-system Schr{\"o}dinger's equation reduces to:
\begin{align}
i \partial_t \ket{ \psi_n^{(j)} (t) } 
& = \big[ H_S (t) - i \Gamma \big] \ket{ \psi_n^{(j)} (t) } 
\nonumber\\
& \quad + \lambda \exp[- i \varepsilon_j (t-t_0)] \ket{ \chi_{n} }.
\label{OSE-2}
\end{align}
The above equation is remarkably simple. It is a simple non-Hermitian version of the Schr{\"o}dinger equation in which the system Hamiltonian is dressed by a constant imaginary part ``$-i \Gamma$'' which captures the decay into the bath. 
Many recent studies of open quantum systems have used non-Hermitian Schr{\"o}dinger equations that only include the first line of Eq.~(\ref{OSE-2}).
However, we see that the influence of the bath is not merely to induce decay, but it also produces the second term that acts a source and makes the equation inhomogeneous.
The balance of these two terms is what allows the existence of non-trivial late time steady states (see Fig.~\ref{fig-system-bath} for depiction).

\subsection{Ideal fermionic bath}
To illustrate that our bath leads to the expected equilibrium when the system is not driven in time, we first consider the the special case in which $H_S (t)$ is time independent,
\begin{align}
H_S (t) \to H_0 = \sum_n \epsilon_n \ket{\chi_n}\bra{\chi_n} ,
\end{align}
Eq.~(\ref{OSE-2}) can be equivalently expressed as
\begin{align}
i\partial_t s_n^{(j)} = [\epsilon_n - i \Gamma ] s_n^{(j)} + \lambda \exp [ -i \varepsilon_j (t-t_0) ] ,
\label{OSE-3}
\end{align}
where
\begin{align}
s_n^{(j)} = \braket{\chi_n |\psi_n^{(j)} (t)} ,
\label{s_n_j}
\end{align}
is the amplitude for the system state $\ket{\chi_n}$.
Solving the above Eq.~(\ref{OSE-3}) gives
\begin{align}
s_n^{(j)} & = - i \lambda 
\exp \Big[ -i \int_{t_0}^t {\text d} t' \, ( \epsilon_n - i \Gamma ) \Big]
\nonumber\\
& \times
\int_{t_0}^t {\text d} t'
\exp \Big[ i \int_{t_0}^{t'} {\text d} t'' (\epsilon_n- i \Gamma - \varepsilon_j ) \Big]
\nonumber\\
& = \frac{\lambda}{\epsilon_n- i \Gamma - \varepsilon_j}
\Big[e^{-(\Gamma + i \epsilon_n) (t-t_0)} - e^{-i \varepsilon_j (t-t_0)}\Big] .
{\label{sn-time-independent}}
\end{align}
Then using Eq.~(\ref{System-Occupation-0}), we obtain the steady state, diagonal density matrix for the system:
\begin{align}
\rho_S(t \to + \infty) = \sum_n f_\Gamma (\epsilon_n) \ket{\chi_n} \bra{\chi_n},
\end{align}
in which $f_\Gamma (\epsilon_n) = \lim_{t\to+\infty} \sum_j f_0 (\varepsilon_j) | s_n^{(j)} |^2$ and reads explicitly as
\begin{align}
& f_\Gamma (\epsilon_n) = \sum_{j} f_0 (\varepsilon_j) \frac{\lambda^2}{(\epsilon_n- i \Gamma - \varepsilon_j)(\epsilon_n + i \Gamma - \varepsilon_j)}
\nonumber\\
= & \int_{-\infty}^{+\infty} {\text d} \omega_b \, f_0 (\omega_b) \frac{\lambda^2 \sum_{j} \delta(\omega_b - \varepsilon_j)   }{(\epsilon_n - i \Gamma - \omega_b)(\epsilon_n + i \Gamma - \omega_b)}
\nonumber\\
= & \int_{-\infty}^{+\infty} \frac{{\text d} \omega_b}{\pi} f_0 (\omega_b) \frac{\Gamma}{(\epsilon_n - \omega_b)^2 + \Gamma^2} ,
\label{f_Gamma_0}
\end{align}
where we used Eqs.~(\ref{broad-flat-bath}) and (\ref{Gamma-definition}) in obtaining the last equation.
The above distribution $f_\Gamma (\epsilon_n)$ shows that when $H_S(t)$ is time independent, the system ``thermalizes'' by approaching a time independent steady state dictated by the initial condition of the bath, $f_0(\omega_b)$, while a finite $\Gamma$ accounts for the broadening of the energy levels of the system due to its coupling to the bath.

Importantly, taking the limit in which the coupling to the bath vanishes from Eq.~(\ref{f_Gamma_0}), we obtain
\begin{align}
\lim_{\Gamma \to 0} f_\Gamma (\epsilon_n) = f_0 (\epsilon_n) ,
\label{f_Gamma_1}
\end{align}
i.e., $f_\Gamma (\epsilon_n)$ reduces to the ideal Fermi-Dirac distribution in the limit of $\Gamma \to 0$.
We will then call this $\Gamma \to 0$ limit of the ``featureless fermionic bath'' an ``ideal fermionic bath''.
The fact that the ideal Fermi-Dirac distribution appears only when the coupling to the bath is vanishingly weak is consistent with general considerations of statistical physics.

However, Eq.~(\ref{f_Gamma_0}) still allow us to obtain analytically the modified occupation at finite coupling to the bath, which will be used in subsequent manipulations.
By integrating over $\omega_b$ in Eq.~(\ref{f_Gamma_0}) using Cauchy's residue theorem, we find that:
\begin{align}
f_\Gamma (\epsilon) = \frac{1}{2} \Big[ f_+ (\epsilon) + f_- (\epsilon) \Big] ,
\label{f_Gamma_2}
\end{align}
where $f_+ (\epsilon) = [ f_- (\epsilon) ]^*$ and they are given by:
\begin{align}
f_\pm (\epsilon) = \frac{1}{2} \pm \frac{i}{\pi}\Psi^{(0)}
\left(\frac{1}{2} \pm i\beta \frac{\epsilon \mp i \Gamma -\mu}{2\pi}\right),
\label{fpm}
\end{align}
with $\Psi^{(0)}$ the 0-th order Polygamma function (or the digamma function). $f_\pm (\epsilon)$ will also appear repeatedly in more general cases.

\subsection{Diagonal and time-periodic Hamiltonians}
\label{DP-occupation}

\subsubsection{Diagonal system Hamiltonian}


In this work, we will develop the above general formalism to the special case where the system Hamiltonian $H_S (t)$ is time dependent but diagonal in the system states. Let us then take the following form for the system Hamiltonian:
\begin{align}
\braket{\chi_n | H_S(t) |\chi_m} = \delta_{nm} [ \epsilon_n + V_n (t) ] = \delta_{nm} \epsilon_n (t) .
\label{H_D}
\end{align}
With this, Eq.~(\ref{OSE-2}) then reduces to
\begin{align}
i\partial_t s_n^{(j)} = [\epsilon_n (t) - i \Gamma ] s_n^{(j)} + \lambda \exp [ -i \varepsilon_j (t-t_0) ] .
\label{OSE-5}
\end{align}
Solving the above Eq.~(\ref{OSE-5}) gives
\begin{align}
s_n^{(j)} (t) & = - i \lambda 
\exp \Big( -i \int_{t_0}^t {\text d} t' \, [ \epsilon_n (t') - i \Gamma ] \Big)
\nonumber\\
& \times
\int_{t_0}^t {\text d} t'
\exp \Big( i \int_{t_0}^{t'} {\text d} t'' [\epsilon_n (t'')- i \Gamma - \varepsilon_j] \Big) ,
{\label{s_n_periodic}}
\end{align}
and then with Eq.~(\ref{System-Occupation-0}), we obtain the diagonal density matrix for the system:
\begin{align}
& \rho_S(t) = \sum_n p_n (t) \ket{\chi_n} \bra{\chi_n},
\\
& p_n (t) = \sum_{j} f_0 (\varepsilon_j) | s_n^{(j)} (t) |^2 .
\label{p_n_0}
\end{align}

\subsubsection{Periodic system Hamiltonian}

Now we consider a periodically driven system. Namely, we take the diagonal elements of the Hamiltonian to be periodic in time:
\begin{align}
\epsilon_n (t+T) = \epsilon_n (t)=
\sum_{l=-\infty}^{+\infty} \epsilon_n^{(l)} \exp[-i l\omega(t-t_0)] ,
\end{align}
where $T$ is the period and $\omega = 2\pi/T$ is the frequency, and
\begin{align}
\epsilon_n^{(l)} = \int_0^T \frac{{\text d} t}{T}
\epsilon_n (t) \exp[i l \omega(t-t_0)]
\label{e_n_l}
\end{align}
is the $l$-th Fourier coefficient for $\epsilon_n (t)$. In particular,
\begin{align}
\bar{\epsilon}_n \equiv \epsilon_n^{(0)} =  \int_{0}^T \frac{{\text d} t}{T} \epsilon_n (t) ,
\label{bepsilon}
\end{align}
is the time-average of the diagonal element of the Hamiltonian, which as we will show next, coincides with the Floquet energy of state $n$.
To see this, notice that the wavefunction that would solve the system Schr{\"o}dinger's equation in the absence of the bath, can be expressed as follows:
\begin{align}
& \exp\Big[-i \int_{t_0}^{t} {\text d} t' \, \epsilon_n (t') \Big] 
\nonumber\\
& = \exp\Big(-i \int_{t_0}^{t} {\text d} t' [\epsilon_n (t') - \bar{\epsilon}_n  ]  \Big)
\times \exp\Big(-i \int_{t_0}^{t}\bar{\epsilon}_n   \Big) 
\nonumber\\
& \equiv \phi_n (t) \times \exp \big[-i \bar{\epsilon}_n (t-t_0)  \big] .
\end{align}
The periodicity of the first factor denoted by $\phi_n (t)$ can be shown explicitly:
\begin{align}
\phi_n (t+T) & = \phi_n (t) \times
\exp\Big(-i \int_{t}^{t+T} {\text d} t' [\epsilon_n (t') - \bar{\epsilon}_n  ]  \Big)
\nonumber\\
& = \phi_n (t) ,
\label{system-wavefunction}
\end{align}
where we used Eq.~(\ref{bepsilon}) in obtaining the second equation.
Therefore we see from second factor in the last line of Eq.~(\ref{system-wavefunction}), that the time-average of the diagonal element of the Hamiltonian is the Floquet energy itself.

Let us now consider the Fourier expansion of the periodic part of the Floquet wavefunction:
\begin{align}
\phi_n (t)
& = \exp\Big(-i \int_{t_0}^{t} {\text d} t' [\epsilon_n (t') - \bar{\epsilon}_n  ]  \Big) 
\nonumber\\
& = \sum_{l = -\infty}^{+\infty} \phi_n^{(l)} \exp [-i l \omega (t -t_0)] ,
\label{phi_nl_1}
\end{align}
or equivalently,
\begin{align}
\phi_n^{(l)} = \frac{1}{T}\int_{t_0}^{t_0 + T} & {\text d} t \bigg[ \exp [ i l \omega (t -t_0) ]
\nonumber\\
& \times \exp \Big( - i \int_{t_0}^{t} {\text d} t' [\epsilon_n (t') - \bar{\epsilon}_n  ] \Big) \bigg].
\label{phi_nl_2}
\end{align}
The above expression makes clear that the amplitude of the harmonics of the wavefunction, $\phi_n^{(l)}$, are functions of the full time dependence of the instantaneous energy $\epsilon_n (t)$, and are independent of the Floquet energy $\bar{\epsilon}_n$.
This property will be crucial later on for purposes of understanding why there is in-gap rectification. In other words, Eq.~(\ref{phi_nl_2}) defines $\phi_n^{(l)}$ as a function of all the harmonics of the time dependent energy from Eq.~(\ref{e_n_l}) as follows:
\begin{align}
\phi_n^{(l)} =
\phi_n^{(l)} (\epsilon_n^{(\pm 1)}, \epsilon_n^{(\pm 2)}, \cdots) .
\label{phi_nl_3} 
\end{align}
Also from Eq.~(\ref{phi_nl_1}) it can be shown that these amplitudes satisfy the following normalization condition:
\begin{align}
\sum_{l=-\infty}^{+\infty} \big| \phi_n^{(l)} \big|^2 =1 .
\end{align}

With Eqs.~(\ref{s_n_periodic}), (\ref{p_n_0}), (\ref{phi_nl_1}), and (\ref{broad-flat-bath}), and by taking the late-time limit that allows to neglect transient terms of the form $\exp[-\Gamma (t-t_0)] \to 0$, we obtain the system steady state occupation:
\begin{align}
p_n (t) & =  \int_{-\infty}^{+\infty} \frac{{\text d} \omega_b}{\pi} f_0(\omega_b)
\nonumber\\
& \quad \times \Gamma \left|\sum_{l=-\infty}^{+\infty} \phi_n^{(l)} \frac{\exp [ - i l \omega(t-t_0) ]}{ \bar{\epsilon}_n -\omega_b - l \omega-i\Gamma}\right|^2 .
\label{p_n_1}
\end{align}
Similar to Eq.~(\ref{f_Gamma_0}), by integrating over $\omega_b$ in Eq.~(\ref{p_n_1}), we find that:
\begin{align}
p_n (t) & = 
\sum_{l,m=-\infty}^{+\infty}
\big[\phi_n^{(m)}\big]^* \phi_n^{(l)} 
\exp\big[ i( m - l ) \omega(t-t_0)\big]
\nonumber\\
& \times
\frac{\Gamma}{2\Gamma+i(m-l)\omega} \Big[ f_+( \bar{\epsilon}_n - l \omega) + f_- ( \bar{\epsilon}_n - m \omega) \Big],
\label{p_n_2}
\end{align}
where $f_{\pm} (\epsilon)$ is given in Eq.~(\ref{fpm}).
The Eq.~(\ref{p_n_2}) is one of the central formulas of our work because it allows to compute expectation values of any equal-time system observables, even at a finite coupling $\Gamma$ to featureless fermionic bath.

The expression in Eq.~(\ref{p_n_2}) captures the steady state occupation of the $n$-th state in the case of featureless fermionic bath, and thus it replaces what would be the Fermi-Dirac distribution in equilibrium.
One important feature of this steady state is that it displays ``synchronization'', namely, it is strictly periodic in the drive:
\begin{align}
p_n (t+T)=p_n (t) .
\end{align}
Remarkably, in the limit of an ``ideal bath'' ($\Gamma \to 0$) the above distribution becomes time independent and it is given by:
\begin{align}
\lim_{\Gamma \to 0} p_n = \sum_{l=-\infty}^{+\infty} |\phi_n^{(l)}|^2 f_0 ( \bar{\epsilon}_n - l\omega) .
\label{p_n_Gamma0}
\end{align}
Here $\bar{\epsilon}_n$ is the Floquet energy of $n$-th state, and $\phi_n^{(l)}$ are the Harmonics of the periodic part of the wave-functions defined in Eq.~(\ref{phi_nl_2}). The reader is encouraged to contrast this occupation function with that in Eq.~(\ref{f_Gamma_1}) obtained when the Hamiltonian was time independent. Notice also that because the occupation function becomes time independent in this limit, there are no time fluctuations of the average fermion occupation of each state $n$.

Thus the distribution is an infinite sum of several Fermi-Dirac distributions with chemical potentials shifted by the various harmonics of the driving frequency $l \omega$ and weighed by amplitudes of the harmonics of the Floquet wavefunctions $|\phi_n^{(l)}|^2$.
It is therefore clear that the occupation of the state is completely different from how the state is filled in equilibrium [see Fig.~\ref{fig-ladder}(a) for an illustration of the non-equilibrium occupation function]. 
One recovers an occupation similar to equilibrium when one neglects all the higher harmonics of $\phi_n^{(l)}$ with $l\neq 0$ and forces by hand the amplitude of the $l=0$ term to be $\phi_n^{(0)} \to 1$, but this is not justified in general (not even perturbatively as we will illustrate in Sec.~\ref{PerturbativeResults}).
We note that the idea that Floquet states are not filled in the same way as equilibrium states has been emphasized in several studies, by using a variety of models for the relaxation when the system is coupled to a heat bath~\cite{genske2015floquet,bilitewski2015scattering,shirai2015condition,seetharam2015controlled,esin2018quantized,seetharam2019steady}.
In fact the expression for the non-equilibrium time independent steady states we find in Eq.~(\ref{p_n_Gamma0}) has been reported before, and is in particular the same kind of expression shown in Eq.~(12) of Ref.~\cite{seetharam2015controlled}.

\subsubsection{Harmonic time dependent driving}

Computing analytically the integral in Eq.~(\ref{phi_nl_2}) that relates the harmonics of the Floquet wavefunction to the harmonics of the energy is in general involved.
There is a simple case where these integrals can be computed in a simple closed analytical form, which is when the time dependent part $V_n (t)$ of the Hamiltonian has a single harmonic:
\begin{align}
V_n (t) = V_n \cos [\omega (t-t_0) ].
\label{V_t_0}
\end{align}
In this case the coefficients $\phi_n^{(l)}$ from Eq.~(\ref{phi_nl_2}) correspond to the $l$-th Bessel function:
\begin{align}
\phi^{(l)}_n =
J_l \big( V_n / \omega \big).
\label{phin_nonp}
\end{align}
Substitution of Eq.~(\ref{phin_nonp}) into Eq.~(\ref{p_n_2}) leads to the following non-perturbative expression for the occupation of the states in the limit of $\Gamma \to 0$:
\begin{align}
\lim_{\Gamma \to 0} p_n = \sum_{l=-\infty}^{+\infty}
J^2_l \big( V_n / \omega \big) f_0 ( \bar{\epsilon}_n - l\omega).
\label{p_n_average}
\end{align}
We therefore see that the occupation in the case of the ideal fermionic bath becomes a sum of several Fermi-Dirac distributions boosted by the different harmonics of the Floquet quasi-energies $ \bar{\epsilon}_n - l \omega$~($l \in {\mathbb Z}$).
It is interesting to note that this ladder-like behavior is analogous to the Tien-Gordon effect that arises in nanostructures that are simultaneously subjected to AC and DC drives~\cite{nazarov2009quantum}.
Similarly as in that case, the ladder behaviour becomes more pronounced as the driving becomes stronger [see Fig.~\ref{fig-ladder}(a)].

\section{Single band model under monochromatic light}
\label{section-2}

In this section we will use the formalism developed in the previous ones to determine the self-consistent occupation of an electronic band driven by an oscillating electric field and demonstrate the existence of in-gap rectification.
Because we are primarily interested here in proving and clarifying the origin of in-gap rectification, we will focus on a simple model of a Bloch band that has vanishing Berry connections.
These bands can display however the in-gap Jerk current effect that arises from the energy band dispersions~\cite{shi2022berry}.
However, other mechanisms driven by the Berry phases, such as the non-linear Hall effect, can also lead to in-gap rectification as we have recently demonstrated~\cite{shi2022berry}.

Let us now consider our system Hamiltonian to be a tight-binding model with a single site per unit cell and a trivial single Bloch band (with no Berry connections) coupled to a uniform monochromatic electric field.
The time dependent system Hamiltonian is:
\begin{align}
& H_S (t) = \int_{\bf k} \epsilon_{\bf k} (t) \ket{\chi_{\bf k}} \bra{\chi_{\bf k}} ,
\\
& \epsilon_{\bf k} (t) \equiv \epsilon ({\bf k} - {\bf A} (t) ) ,
\quad
\int_{\bf k} \equiv \int_\text{BZ} \frac{{\text d} {\bf k}}{(2\pi)^d} .
\label{HS-1band}
\end{align}
The system states are now labelled by the wave vector ${\bf k}$ and $\epsilon ({\bf k})$ is the unperturbed band dispersion.
We assume a monochromatic electric field which leads to the periodic vector potential using ${\bf E} (t) = - \partial_t {\bf A}(t)$:
\begin{align}
{\bf A} (t) = -\frac{i}{\omega} {\bf E}_\omega \exp (-i \omega t) + \text{c. c.}
\label{AfromE}
\end{align}

\subsection{Electric current in the steady state}
Since the system Hamiltonian is diagonal in crystal momenta ${\bf k}$, we can apply the formalism of Sec.~\ref{DP-occupation} to compute the steady state occupation of each momenta ${\bf k}$, by replacing the label in previous sections $n \to {\bf k}$.
If we denote the occupation of each state by $p_{\bf k}(t)$, then the system's electric current reads as follows:
\begin{align}
{\bf j} (t) & =
\int_{\bf k} p_{\bf k} (t)
\nabla_{\bf k} \epsilon_{\bf k} (t)
\nonumber\\
& = 
\sum_{s=-\infty}^{+\infty} {\bf j}^{(s)} \exp[- i s \omega(t-t_0)],
\label{current}
\end{align}
where we set $e = \hbar = 1$ throughout the paper.
By combining Eqs.~(\ref{e_n_l}) and (\ref{p_n_2}), the weight of each oscillating mode of the electric current can be written as:
\begin{align}
& {\bf j}^{(s)} =
\int_{\bf k}
\sum_{m,l=-\infty}^{+\infty}
\frac{\Gamma}{2\Gamma+i(l - s)\omega}\big[ \phi_{\bf k}^{(m)} \big]^*  \phi_{\bf k}^{(s + m - l)} \nonumber\\
& \times \big[ f_+( \bar{\epsilon}_{\bf k} - (s + m - l)\omega)+f_-( \bar{\epsilon}_{\bf k} - m \omega)\big] \nabla_{\bf k} \epsilon_{\bf k}^{(l)}.
\label{currentgeneral}
\end{align}

Interestingly, as discussed in Sec.~\ref{DP-occupation}, in the limit of an ideal heat bath $\Gamma \to 0$, the distribution function $p_{\bf k}(t)$ becomes time independent, and therefore the time averaged electric current (also referred to as rectified current), is given by:
\begin{align}
& \bar{\bf j} = \int_0^T \frac{{\text d} t}{T} {\bf j} (t) = 
\int_{\bf k}
p_{\bf k}
\nabla_{\bf k} \bar{\epsilon}_{\bf k} ,
\label{cleanrectification}
\end{align}
where $\bar{\epsilon}_{\bf k}$ is the Floquet energy of the band and in our current simple single-band model, and is given by the time averaged band energy ($l=0$ component):
\begin{align}
\bar{\epsilon}_{\bf k} \equiv \epsilon_{\bf k}^{(0)} = \int_{0}^T \frac{{\text d} t}{T} \epsilon ({\bf k}-{\bf A}(t)) .
\end{align}
Therefore, we see that Eq.~(\ref{cleanrectification}) has a resemblance to how one would compute the current in a time independent equilibrium system, but with the equilibrium Fermi-Dirac distribution replaced by occupation function $p_{\bf k}$, and the bare band dispersion replaced by the dressed Floquet band energy $\bar{\epsilon}_{\bf k}$.
At first glance, this point of view might suggest that the time averaged rectified current vanishes in the ideal limit of $\omega \gg \Gamma \to 0$, just in the same way it is expected to vanish in a time independent equilibrium system.
In fact, several classic and more recent works have incorrectly taken this point of view that the non-equilibrium steady state occupation $p_{\bf k}$ is a Fermi-Dirac distribution of the dressed Floquet band energy~\cite{belinicher1986transient,ivchenko1988magneto,onishi2022effects,golub2022raman, pershoguba2022direct} (see Appendix~\ref{appendixE} for detailed comments on previous works).
However, as we have shown in Sec.~\ref{DP-occupation}, the correct occupation of the states in the non-equilibrium steady state is not a simple Fermi-Dirac distribution, but it is given by the following expression [see Eqs.~(\ref{phi_nl_3}) and (\ref{p_n_Gamma0})]:
\begin{align}
p_{\bf k} ( \bar{\epsilon}_{\bf k},\epsilon_{\bf k}^{(\pm1)},\cdots) \equiv
\sum_{l=-\infty}^{+\infty} |\phi_{\bf k}^{(l)}|^2 f_0 ( \bar{\epsilon}_{\bf k} - l\omega) .
\label{p_k_Gamma0}
\end{align}
In the argument of $p_{\bf k}$ in the above expression, we have emphasized that $p_{\bf k}$ is not only a function of the Floquet band energy $\bar{\epsilon}_{\bf k}$, but also of all the higher harmonics $\epsilon_{\bf k}^{(\pm1)}, \, \epsilon_{\bf k}^{(\pm2)} \, \cdots$ of the time dependent energy $\epsilon_{\bf k}(t)$ through its dependence on the amplitudes $\phi_{\bf k}^{(l)}$ [see Eqs.~(\ref{phi_nl_2}) and (\ref{phi_nl_3})].
Precisely because of this, the rectification current $\bar{\bf j}$ can not be expressed as an integral of a total derivative over the Brillouin zone and generally does not vanish, i.e.,
\begin{align}
\bar{\bf j} \neq \int_{\bf k} \nabla_{\bf k} \tilde{P} (\bar{\epsilon}_{\bf k}) = 0 , 
\label{j_neq_0}
\end{align}
where $\tilde{P} (\bar{\epsilon}_{\bf k})$ would be defined through
\begin{align}
\frac{ \partial \tilde{P} (\bar{\epsilon}_{\bf k}) }{ \partial \bar{\epsilon}_{\bf k} } \equiv \tilde{p}_{\bf k} (\bar{\epsilon}_{\bf k}) ,
\label{P_k}
\end{align}
which would be possible if the occupation depended only on the dressed Floquet energy $p_{\bf k} \to \tilde{p}_{\bf k} ( \bar{\epsilon}_{\bf k} )$  [but this is not the case for Eq.~(\ref{j_neq_0})].

Therefore, we see that in general a non-zero rectified current is expected in the non-equilibrium steady state, even in the limit of the $\omega \gg \Gamma \to 0$. 
As we will show in detail in the following section, this finite rectified current remains non-zero within the optical gap of a metal, even within the usual second order of perturbation theory in the amplitude of the electric field for which rectification currents are typically computed.
These findings further substantiate our recent work showing the existence of in-gap rectification~\cite{shi2022berry} but appear in tension with some other statements in the literature~\cite{de2020difference,onishi2022effects,golub2022raman,belinicher1986transient,ivchenko1988magneto,pershoguba2022direct}.
In Appendix~\ref{appendixE}, we comment in more detail on some of these other works clarifying some partial agreements but also pointing out some of their imprecisions and incorrect statements.

\subsection{Perturbative results}
\label{PerturbativeResults}

In this subsection we will compute perturbatively the electric current in powers of electric field to the currents at modes [see Eqs.~(\ref{current}) and (\ref{currentgeneral})]: $s = 0$ representing rectification conductivity, $s = 1$ representing linear conductivity.
$s = 2$ representing second harmonic generation is discussed in Appendix~\ref{appendixA}.
We will show explicitly that even to 2nd order in electric fields, the non-equilibrium distribution in the steady state for an ideal bath, $p_{\bf k}$, differs clearly from the naive Fermi-Dirac distribution evaluated in the dressed Floquet bands.
This will allow us to compute analytically the rectification conductivities and prove rigorously that they remain finite within the optical gap of the metal.

Although our conclusions and formulae are valid and can be used for any single band model (with no Berry connections) in arbitrary dimensions, 
for simplicity we will illustrate our results for a simple 1D model with the following band dispersion:
\begin{align}
\epsilon (k_x) = -t_1 \cos (a_0 k_x)-t_2\sin(2 a_0 k_x) + \epsilon_0 ,
\label{HammOneBand}
\end{align}
where $\epsilon_0$ is a constant that we have added for convenience in order to shift the band energy so that it lies within $0$ and $\Delta$ [See Fig.~{\ref{fig-dispersion-conductivity}}(b)], and $a_0$ is the lattice constant.
Notice that the above band-structure breaks not only inversion, which is always needed to have rectification, but also time-reversal symmetry, and therefore it has no symmetry relating ${\bf k} \to -{\bf k}$. As we will see, this is indeed crucial in order to obtain a non-zero in-gap rectification conductivities for the models without Berry curvature that we are considering in this study.
More generally, as discussed in Ref.~\cite{shi2022berry}, in the case of bands with non-trivial Berry connections one can alternatively obtain a non-zero in-gap rectification, e.g., via the Berry-Dipole effect by breaking time reversal symmetry only by having a circularly polarized light instead of having a time-reversal breaking band-structure.

\subsubsection{Occupation function to the second order of electric field}

We begin by deriving the explicit perturbative expressions for $\epsilon_{\bf k}^{(l)}$ and $\phi_{\bf k}^{(l)}$ discussed in the previous sections and can be computed from Eqs.~(\ref{e_n_l}) and (\ref{phi_nl_2}) by replacing $n \to {\bf k}$.
Up to the second order in the electric field, it is sufficient to expand the band dispersion up to the same second order, namely:
\begin{align}
& \epsilon({\bf k}-{\bf A}(t)) 
=
\bar{\epsilon}_{\bf k} +\epsilon_{\bf k}^{(1)} e^{-i\omega(t-t_0)}
+
\epsilon_{\bf k}^{(-1)} e^{i\omega(t-t_0)}
\nonumber\\
&\qquad +
\epsilon_{\bf k}^{(2)} e^{-2i\omega(t-t_0)}
+
\epsilon_{\bf k}^{(-2)}e^{2i\omega(t-t_0)}
+ \cdots ,
\end{align}
Using Eq.~(\ref{AfromE}), this perturbative expansion leads to the following expressions for $\epsilon_{\bf k}^{(l)}$:
\begin{align}
& \bar{\epsilon}_{\bf k} \equiv  \epsilon_{\bf k}^{(0)} =  \epsilon ({\bf k}) +
\frac{1}{\omega^2}
\sum_{\alpha \beta}
\partial_\alpha\partial_\beta\epsilon({\bf k}) \, E^\alpha_{\omega} E^{\beta}_{-\omega} + O(|E_\omega|^4) ,
\nonumber\\
& \epsilon_{\bf k}^{(1)} = 
\frac{i}{\omega}
\sum_\alpha
\partial_\alpha \epsilon ({\bf k}) \, E_\omega^\alpha
+ O(|E_\omega|^3) ,
\nonumber\\
& \epsilon_{\bf k}^{(2)} = -\frac{1}{2 \omega^2}
\sum_{\alpha \beta}
\partial_\alpha \partial_\beta \epsilon({\bf k}) \, E_\omega^\alpha E_\omega^\beta + O(|E_\omega|^4) ,
\nonumber\\
& \epsilon_{\bf k}^{(-l)} = \big[ \epsilon_{\bf k}^{(l)} \big]^* .
\label{eexp}
\end{align}
We can use Eq.~(\ref{phi_nl_2}) to perturbatively evaluate $\phi_{\bf k}^{(l)}$ leading to:
\begin{align}
& \phi_{\bf k}^{(0)} = 1 
-
\frac{\epsilon_{\bf k}^{(1)}-\epsilon_{\bf k}^{(-1)}}{\omega}
\nonumber\\
&\quad +\frac{\big[\epsilon_{\bf k}^{(1)}\big]^2 + \big[\epsilon_{\bf k}^{(-1)}\big]^2 - 4 \epsilon_{\bf k}^{(1)} \epsilon_{\bf k}^{(-1)} - \epsilon_{\bf k}^{(2)} + \epsilon_{\bf k}^{(-2)} }{2\omega^2},
\nonumber\\
& \phi_{\bf k}^{(1)} = -\frac{\epsilon_{\bf k}^{(1)}}{\omega}
-\frac{\epsilon_{\bf k}^{(1)}
\big[ \epsilon_{\bf k}^{(1)}-\epsilon_{\bf k}^{(-1)} \big]}{\omega^2},
\nonumber\\
& \phi_{\bf k}^{(-1)} = \frac{\epsilon_{\bf k}^{(-1)}}{\omega}
-
\frac{\epsilon_{\bf k}^{(-1)}
\big[ \epsilon_{\bf k}^{(-1)}-\epsilon_{\bf k}^{(1)} \big]}{\omega^2},
\nonumber\\
& \phi_{\bf k}^{(2)} = \frac{\big[\epsilon_{\bf k}^{(1)}\big]^2-\epsilon_{\bf k}^{(2)}}{2\omega^2},
\quad
\phi_{\bf k}^{(-2)} = \frac{\big[\epsilon_{\bf k}^{(-1)}\big]^2+\epsilon_{\bf k}^{(-2)}}{2\omega^2}.
\end{align}
The other $\phi_{\bf k}^{(l)}$ with $|l| > 2$ will scale with higher powers of electric fields, and therefore can be neglected to second order.
The norm squared of those terms above are:
\begin{align}
& \big|\phi_{\bf k}^{(0)}\big|^2 = 1 - \frac{2\big|\epsilon_{\bf k}^{(1)}\big|^2}{\omega^2}+O(|E_\omega|^3),
\nonumber\\
& \big|\phi_{\bf k}^{(1)}\big|^2 =
\frac{\big|\epsilon_{\bf k}^{(1)}\big|^2}{\omega^2}+O(|E_\omega|^3),
\nonumber\\
& \big|\phi_{\bf k}^{(2)}\big|^2 =O(|E_\omega|^4) .
\label{squares}
\end{align}
Therefore the ideal occupation function $p_{\bf k}$ in the limit $\Gamma \to 0$ to second order in electric fields reads as
\begin{align}
p_{\bf k} & = \Big( 1 - \frac{2\big|\epsilon_{\bf k}^{(1)}\big|^2}{\omega^2} \Big)f_0 ( \bar{\epsilon}_{\bf k})
\nonumber\\
& + \frac{\big|\epsilon_{\bf k}^{(1)}\big|^2}{\omega^2} f_0 ( \bar{\epsilon}_{\bf k} - \omega) 
+ \frac{\big|\epsilon_{\bf k}^{(-1)}\big|^2}{\omega^2} f_0 ( \bar{\epsilon}_{\bf k} + \omega) .
\label{ideal-occupation-2nd}
\end{align}
The above expansion contains all the correct terms to second order in electric fields, even though it is not strictly perturbative, because the Floquet band energy $\bar{\epsilon}_{\bf k}$ also includes implicitly a correction of order $|E_\omega|^2$ [see Eq.~(\ref{eexp})].
In other words, if one wants to obtain a strictly perturbative expansion to order $|E_\omega|^2$ one simply needs to Taylor expand the Fermi-Dirac distribution $f_0(\bar{\epsilon}_{\bf k})$ above as well.
However we find it convenient to keep the above form, with the understanding that we can only trust its predictions to order $|E_\omega|^2$.

Let us now comment on the significance of Eq.~(\ref{ideal-occupation-2nd}).
We see above that even to second order, the non-equilibrium distribution, $p_{\bf k}$, contains not only the Fermi-Dirac distribution evaluated for the Floquet bands, $f_0(\bar{\epsilon}_{\bf k})$, but also several other terms that make it clearly deviate from $f_0(\bar{\epsilon}_{\bf k})$.
As we will see these additional terms, are precisely the ones that lead to a finite in-gap rectification in the clean limit $\Gamma \to 0$.
In Appendix~\ref{appendixD}, we also demonstrate that the above occupation function agrees with the one obtained from a simpler Boltzmann/relaxation-time description in the limit $\omega \ll \bar{\epsilon}_{\bf k}$. Notice also that the above occupation differs even to up second order $|E_\omega|^2$ from the naive Fermi-Dirac occupation  of the Floquet band, $f_0(\bar{\epsilon}_{\bf k})$, that was pressumed in Refs.~\cite{onishi2022effects,golub2022raman,belinicher1986transient,ivchenko1988magneto,pershoguba2022direct} (see Appendix Appendix~\ref{appendixE} for further comments on previous studies).

\subsubsection{Linear conductivity}
The linear conductivity is defined from:
\begin{align}
j_\alpha^{(1)} = \sigma_\Gamma^{\alpha\beta} (\omega) E^\beta_\omega + O (|E_\omega|^3) ,
\end{align}
where the sub-index $\Gamma$ emphasizes a finite coupling of the system to the bath.
Using Eqs.~(\ref{currentgeneral}), (\ref{phi_nl_2}), (\ref{e_n_l}), the exact conductivity of our model at finite coupling to the bath is found to be: 
\begin{align}
\sigma_\Gamma^{\alpha\beta}(\omega) = \frac{i}{\omega} & \int_{\bf k}  f_{\Gamma}(\bar{\epsilon}_{\bf k}) \partial_\alpha \partial_\beta \bar{\epsilon}_{\bf k}
\nonumber\\
& + \int_{\bf k}
\frac{\partial_\alpha \bar{\epsilon}_{\bf k} \partial_\beta\bar{\epsilon}_{\bf k}}{\omega^2}\frac{i\Gamma}{2\Gamma-i\omega}{\cal L}_1(\bar{\epsilon}_{\bf k},\omega),
\label{sigma1full}
\end{align}
where $\partial_\gamma \equiv \partial/\partial k_\gamma$, and
\begin{align}
{\cal L}_1 (\bar{\epsilon}_{\bf k},\omega) = f_+(\bar{\epsilon}_{\bf k}) & + f_-(\bar{\epsilon}_{\bf k}+\omega)
\nonumber\\
& - f_+(\bar{\epsilon}_{\bf k}-\omega) - f_-(\bar{\epsilon}_{\bf k}) ,
\end{align}
where $f_{\pm}$ are defined in Eq.~(\ref{fpm}).
Just as for Eq.~(\ref{ideal-occupation-2nd}), we have kept the dressed Floquet band energy, $\bar{\epsilon}_{\bf k}$, in the integrands of Eq.~(\ref{sigma1full}), and therefore this is not a strictly perturbative expression.
But if desired, the strictly perturbative expression can simply be obtained from the one above by replacing the dressed Floquet band energy dispersion by the bare unperturbed band dispersion: $\bar{\epsilon}_{\bf k} \to \epsilon ({\bf k})$.
This also applies to the subsequent formulas of this section.

\begin{figure}[t]
\centering
\includegraphics[width=0.48\textwidth]{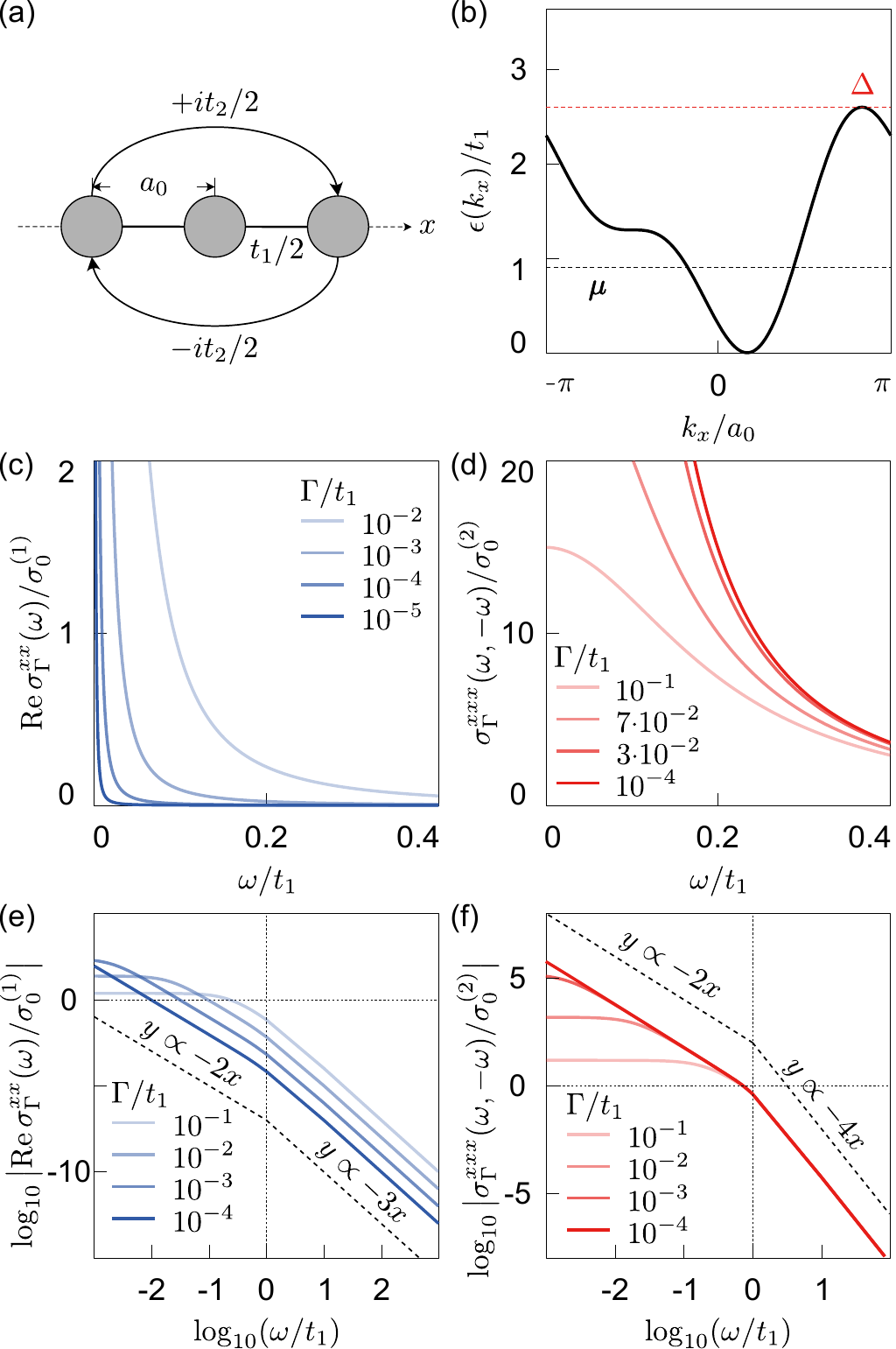}
\caption{(a) The 1D tight binding model whose inversion and time-reversal symmetries are broken by the next-nearest-neighbour hopping $\pm i t_2 / 2$, and its (b) dispersion relation with $0$ the band bottom and $\Delta$ the band top. (c) Real part of the dimensionless linear conductivity $\text{Re}\, \sigma_\Gamma^{xx} (\omega) / \sigma_0^{(1)} $ illustrating how it vanishes at finite frequency as $\Gamma \to 0$ (which defines the optical transparency region), and (d) dimensionless rectification conductivity $\sigma_\Gamma^{xxx} (\omega,-\omega) / \sigma_0^{(2)} $ for different $\Gamma$ illustrating the existence of in-gap rectification in the metal, namely that it approaches a finite non-zero value in the limit of $\Gamma \to 0$ at finite $\omega$.
The characteristic linear and second order conductivities in 1D used here are $\sigma_0^{(1)} = a_0 \cdot e^2 / \hbar$ and $\sigma_0^{(2)} = a_0^2 \tau_0 \cdot e^3 / \hbar^2$ with $\tau_0 = \hbar / t_1$.
(e) and (f) Log-log plots of $\text{Re}\, \sigma_\Gamma^{xx} (\omega) / \sigma_0^{(1)} $ and $\sigma_\Gamma^{xxx} (\omega,-\omega) / \sigma_0^{(2)} $ for different $\Gamma$ illustrating their power dependencies over $\omega$ in different frequency ranges. 
Parameters used: $a_0 = 1$, $t_1 / t_2 = 2$, $\mu = 5 t_1 /7$, $\beta_0 = 10^9 / t_1$.}
\label{fig-dispersion-conductivity}
\end{figure}

In the clean limit ($\omega \neq 0$ and $\Gamma \to 0$), the above expression reduces to the standard Drude form:
\begin{align}
\lim_{\Gamma \to 0}\sigma_\Gamma^{\alpha\beta} (\omega) = \frac{i}{\omega}\int_{\bf k}  f_{0}(\bar{\epsilon}_{\bf k}) \partial_\alpha \partial_\beta  \bar{\epsilon}_{\bf k}.
\label{linear-clean}
\end{align}
Therefore, we see that the real part of the linear conductivity at finite frequency vanishes when $\Gamma \to 0$.
In Fig.~\ref{fig-dispersion-conductivity}(c) we illustrate this in detail for the simple model 1D from Eq.~(\ref{HammOneBand}).
The above Drude form follows from the fact that to the linear order of the electric field, the ideal occupation function $p_{\bf k}$ in the limit $\Gamma \to 0$ is the same with the equilibrium Fermi-Dirac distribution [see Eq.~(\ref{ideal-occupation-2nd})].

In the DC limit $\omega \to 0$ the linear conductivity approaches a finite Drude-like value (see Appendix~\ref{appendixA} for details): 
\begin{align}
\lim_{\omega \to 0}
\sigma_\Gamma^{\alpha\beta} (\omega)
= \frac{1}{2} 
\int_{\bf k} 
& \partial_\alpha\partial_\beta \bar{\epsilon}_{\bf k}
\Big[
\frac{f_\Gamma(\bar{\epsilon}_{\bf k})}{\Gamma}
-\frac{\partial g_\Gamma(\bar{\epsilon}_{\bf k})}{\partial\bar{\epsilon}_{\bf k}}
\Big]
\nonumber\\
\approx \frac{1}{2}\int_{\bf k} 
& \partial_\alpha\partial_\beta \bar{\epsilon}_{\bf k} 
\Big[
\frac{f_0(\bar{\epsilon}_{\bf k})}{\Gamma} 
+ 
O(\Gamma)
\Big] ,
\end{align}
in which
\begin{align}
g_\Gamma (\epsilon) = \frac{1}{2i} \Big[ f_+ (\epsilon) - f_- (\epsilon) \Big] 
\end{align}
is the imaginary part of $f_+ (\epsilon)$ defined in Eq.~(\ref{fpm}).
Therefore the clean limit of the DC conductivity resembles the prediction of the classic Drude theory for $\tau \equiv 1/(2\Gamma)$, and has a Drude peak in the DC limit when the chemical potential of the bath is within the bandwidth of the system $\mu\in[0,\Delta]$ [see Fig.~\ref{fig-dispersion-conductivity}(c)]. 
The fact that the conductivity is finite when $\omega \to 0$ and has the expected Drude behavior, evidences that our simple bath produces the correct behavior for the relaxation of currents.

In the limit in which the frequency is small compared to the bandwidth but much larger than $\Gamma$,
we obtain the usual decay power $1/\omega^2$ associated with the Drude behavior [see Fig.~\ref{fig-dispersion-conductivity}(e), left-hand side region]:
\begin{align}
\lim_{\Gamma\ll\omega\ll\Delta}{\rm Re}
\big[\sigma_\Gamma^{\alpha\beta}(\omega)
\big]
= -\frac{2\Gamma}{\omega^2}
\int_{\bf k} (\partial_\alpha \bar{\epsilon}_{\bf k})(\partial_\beta \bar{\epsilon}_{\bf k})
\frac{\partial f (\bar{\epsilon}_{\bf k})}{\partial \bar{\epsilon}_{\bf k}} .
\end{align}
On the other hand, in the ultra-large frequency regime when the frequency greatly exceeds even the bandwidth, the real part of the linear conductivity has a different scaling from that of Drude theory:
\begin{align}
\lim_{\omega\gg\Delta}{\rm Re}
\big[\sigma_\Gamma^{\alpha\beta}(\omega)
\big]
= \frac{\Gamma}{\omega^3}
\int_{\bf k} (\partial_\alpha \bar{\epsilon}_{\bf k})(\partial_\beta \bar{\epsilon}_{\bf k}) ,
\end{align}
decaying as $1/\omega^3$ [see Fig.~\ref{fig-dispersion-conductivity}(e), right-hand side region].

\subsubsection{Rectification conductivity}
The rectification conductivity 
is a three-index tensor that relates the time averaged current [namely the average DC current corresponding to $s=0$ in Eq.~(\ref{current})] to the bilinears of electric field amplitudes.
Without loss of generality, we define it by choosing the following symmetry convention for indices of the electric field bilinears:
\begin{align}
j_\gamma^{(0)} & =
\sigma_{\Gamma}^{\gamma\alpha\beta}(\omega,-\omega)E^\alpha_\omega (E^{\beta}_\omega)^*
\nonumber\\
& \qquad + \sigma_{\Gamma}^{\gamma\alpha\beta}(-\omega,\omega) (E_\omega^{\alpha})^* E^\beta_\omega  + O(|E_\omega|^4) .
\end{align}
The exact rectification conductivity of our model at finite coupling to the bath, $\Gamma$, is given by:
\begin{align}
& \sigma^{\gamma\alpha\beta}_{\Gamma}(\omega,-\omega) 
\nonumber\\
& = \int_{\bf k} \frac{\partial_\gamma \bar{\epsilon}_{\bf k} \partial_\alpha\bar{\epsilon}_{\bf k}\partial_\beta\bar{\epsilon}_{\bf k}}{2\omega^4}[f_\Gamma(\bar{\epsilon}_{\bf k}+\omega)+f_\Gamma(\bar{\epsilon}_{\bf k}-\omega)-2f_\Gamma(\bar{\epsilon}_{\bf k})]
\nonumber\\
& \qquad +
\frac{\Gamma}{2\Gamma - i\omega}\int_{\bf k} \frac{\partial_\alpha\bar{\epsilon}_{\bf k}\partial_\gamma\partial_\beta\bar{\epsilon}_{\bf k}}{2\omega^3}\mathcal{L}_1(\bar{\epsilon}_{\bf k},\omega)
\nonumber\\
& \qquad + \frac{\Gamma}{2\Gamma + i\omega}\int_{\bf k} \frac{\partial_\beta\bar{\epsilon}_{\bf k}\partial_\gamma\partial_\alpha\bar{\epsilon}_{\bf k}}{2\omega^3}\mathcal{L}^*_1(\bar{\epsilon}_{\bf k},\omega) .
\label{sigma2full}
\end{align}

The DC limit of the rectification conductivity can be shown to be (see Appendix~\ref{appendixB} for details): 
\begin{align}
& \lim_{\omega \to 0}
\sigma_{\Gamma}^{\gamma\alpha\beta}(\omega,-\omega) 
\nonumber\\
& =\frac{1}{4}
\int_{\bf k}\partial_\alpha\partial_\beta\partial_\gamma \bar{\epsilon}_{\bf k}
\Big[ \frac{f_\Gamma(\bar{\epsilon}_{\bf k})}{\Gamma^2} -\frac{1}{\Gamma}\frac{\partial g_\Gamma(\bar{\epsilon}_{\bf k})}{\partial\bar{\epsilon}_{\bf k}}-\frac{1}{3}\frac{\partial^2 f_\Gamma(\bar{\epsilon}_{\bf k})}{\partial\bar{\epsilon}_{\bf k}^2}\Big]
\nonumber\\
& 
\approx \frac{1}{4}\int_{\bf k} \partial_\alpha\partial_\beta \partial_\gamma \bar{\epsilon}_{\bf k}
\Big[
\frac{f_0(\bar{\epsilon}_{\bf k})}{\Gamma^2}
+
O(\Gamma^0)
\Big] .
\label{DC-rect}
\end{align}
The leading term of the above expression in the second line coincides with the Jerk conductivity predicted within the relaxation time approximation from a simple Boltzmann-relaxation-time formalism~\cite{sodemann2015quantum,matsyshyn2019nonlinear,shi2022berry}.
For an illustration see Fig.~\ref{fig-dispersion-conductivity}(d).
We have also verified that the above $\omega \to 0$ limit of the rectification conductivity is identical to the $\omega \to 0$ limit of the second-harmonic generation conductivity $\sigma_\Gamma^{\gamma \alpha \beta}(\omega,\omega)$
(see Appendix~\ref{appendixC} for details).

Let us now focus on the main regime of our interest, which is the ``clean-limit'' in which the relaxation rate vanishes ($\Gamma \to 0$) while the frequency remains finite. The exact expression for the rectification conductivity in this limit is given by:
\begin{align}
\lim_{\Gamma \to 0}
& \sigma_\Gamma^{\gamma \alpha\beta} (\omega,-\omega)=\frac{1}{2\omega^4}\int_{\bf k} (\partial_\gamma \bar{\epsilon}_{\bf k}) (\partial_\alpha\bar{\epsilon}_{\bf k})(\partial_\beta\bar{\epsilon}_{\bf k})
\nonumber\\
& \times \big[ f_0 (\bar{\epsilon}_{\bf k}+\omega)+f_0 (\bar{\epsilon}_{\bf k}-\omega)-2f_0 (\bar{\epsilon}_{\bf k}) \big]
.
\label{clean-rec}
\end{align} 
Notice that the above rectification conductivity would vanish under any symmetry that enforces $\bar{\epsilon}_{\bf k} = \bar{\epsilon}_{-{\bf k}}$, such as time reversal or inversion symmetry. 
Therefore, the above expression proves one of our central claims, namely that the rectification conductivity remains finite at finite frequency within the optical transparency region of the metal.
The  ``transparency'' here refers to the fact that the real part of the linear conductivity vanishes in this same limit $\omega \gg \Gamma \to 0$. 
We illustrate this behavior in Fig.~\ref{fig-dispersion-conductivity}(d) for our toy 1D model,
confirming that the in gap rectification is possible.
The origin of this finite rectification conductivity can be traced back to the fact that to the second order of the electric field, the ideal occupation function $p_{\bf k}$ in the limit $\Gamma \to 0$ is different from the equilibrium Fermi-Dirac distribution [see Eq.~(\ref{ideal-occupation-2nd})].

\begin{figure}[t]
\centering
\includegraphics[width=0.48\textwidth]{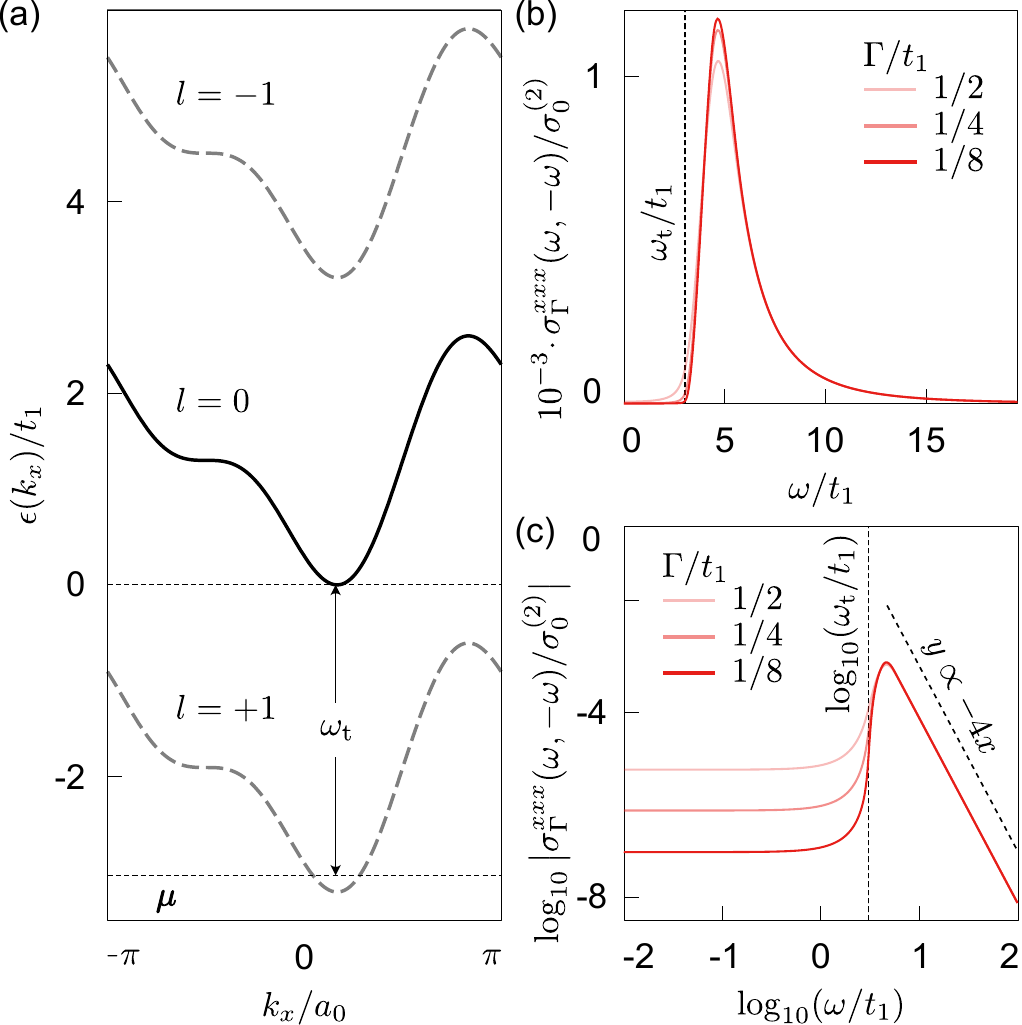}
\caption{(a) Schematic of the original band (denoted by solid line $l=0$) and the boosted Floquet bands (denoted by dashed lines $l=\pm 1$). Here the chemical potential $\mu$ is below the original band. The threshold frequency $\omega_\text{t}$ is the minimum frequency for boosted Floquet bands to cross the chemical potential.  (b) and (c) dimensionless rectification conductivity $\sigma_\Gamma^{xxx} (\omega,-\omega) / \sigma_0^{(2)} $ and its Log-log plots for different $\Gamma$, showing that  rectification conductivity is non-zero when $\omega > \omega_\text{t}$.
Parameters used are the same with those in Fig.~\ref{fig-dispersion-conductivity}.}
\label{fig-threshold}
\end{figure}

While the expression of Eq.~(\ref{clean-rec}) is the exact clean limit of the rectification conductivity in our model, it can be shown that this expression reduces to the more familiar expression for the Jerk current prediction of the simple Boltzmann-relaxation-time expression in the limit in which the frequency is small compared to the bandwidth, namely
$\Gamma \ll \omega \ll\Delta$, and it is given by:
\begin{align}
\lim_{\omega \to 0}
\lim_{\Gamma \to 0}
\sigma_\Gamma^{\gamma\alpha\beta} (\omega,-\omega) =\frac{1}{\omega^2}
\int_{\bf k}  f_0(\bar{\epsilon}_{\bf k}) \partial_\alpha\partial_\beta\partial_\gamma  \bar{\epsilon}_{\bf k} ,
\label{dc-clean-rec}
\end{align}
which coincides with Eq.~(21) of Ref.~\cite{shi2022berry} for the Jerk mechanism which has a $1/\omega^2$ decaying power [see Fig.~\ref{fig-dispersion-conductivity}(f), left-hand side region]. More details of this agreement with the simpler Boltzmann approach are discussed in Appendix~\ref{appendixD}.

Interestingly, in the ``ultra-high'' frequency limit, when the frequency is much larger than the bandwidth $\omega \gg \Delta$, the clean rectification conductivity transits to a different scaling and decays much faster [see Fig.~\ref{fig-dispersion-conductivity}(f), right-hand side region]:
\begin{align}
\lim_{\omega \gg \Delta}
& \lim_{\Gamma \to 0}
\sigma_\Gamma^{\gamma\alpha\beta} (\omega,-\omega)
\nonumber\\
& = \frac{1}{2\omega^4}\int_{\bf k} \big[1-2f_0(\bar{\epsilon}_{\bf k})\big]
(\partial_\alpha\bar{\epsilon}_{\bf k})(\partial_\beta\bar{\epsilon}_{\bf k})(\partial_\gamma\bar{\epsilon}_{\bf k}).
\label{ultrahigh-rectification}
\end{align}
In contrast to the Boltzmann-relaxation-time result where the large frequency regime is controlled by the third momentum derivative of the band dispersion, here, the large frequency response is controlled by the third power of band velocity, which is a different intrinsic property of the band.

It is interesting to note that the expression in Eq.~(\ref{ultrahigh-rectification}) remains finite even when the unperturbed band is either fully occupied [$f_0 (\bar{\epsilon}_{\bf k})= 1$] or fully empty [$f_0 (\bar{\epsilon}_{\bf k})= 0$], namely the system would be nominally an insulator without a Fermi surface. 
This behavior is possible because our bath does not conserve the total particle number of the system, and therefore, there appears a finite occupation of the bands when they are driven by the electric field, even if the bands were initially empty in the distant past before turning on the time dependent drive.
In other words, all our calculations are performed strictly for a bath with fixed chemical potential but not fixed density.
The appearance of a finite occupation of the bands to second order of perturbation theory occurs when the frequency exceeds the threshold so that one of the copies of the Floquet bands boosted by $\pm \omega$ crosses the chemical potential, as depicted in Fig.~\ref{fig-threshold}. 

\section{Summary and discussion}

We have shown rigorously that the occupation of states in a periodically driven fermionic system coupled to a featureless fermionic heat bath approaches a time independent occupation function in the limit in which the coupling to this bath is vanishingly small.
This occupation function can be computed analytically and differs from the naive Fermi-Dirac occupation of the dressed Floquet energies.
This non-equilibrium steady state occupation instead resembles a staircase version of the Fermi-Dirac distribution [see Fig.~\ref{fig-ladder}(a) for an illustration], and also cannot be expressed as a function of the Floquet energy alone, but in general contains information on all the harmonics encoding the full time dependence of the Hamiltonian. 

We applied these results to the case in which the fermionic system has a Hamiltonian corresponding to a single Bloch band without Berry connections (e.g. arising from a tight-binding model with a single site per unit cell) driven by a monochromatic electric field.
We showed that this staircase Fermi-Dirac distribution leads to a finite rectification conductivity within the optical transparency region of a metal, which at small frequencies compared to the bandwidth agrees exactly with the prediction of the Jerk current effect expected from a simpler Boltzmann-relaxation-time description~\cite{matsyshyn2019nonlinear,shi2022berry}.
Because the oscillating electric field is monochromatic, this rectification conductivity does not arise because of the frequency difference effect of Ref.~\cite{de2020difference} or the Raman-like scattering effect of Refs.~\cite{onishi2022effects,golub2022raman}. 

Our results validate our recent findings~\cite{shi2022berry} that in-gap rectification within the optical transparency region of metals are indeed possible, even in the limit in which carrier relaxation rates vanish,
and clarify a 
discussion surrounding this matter~\cite{de2020difference,onishi2022effects,golub2022raman,belinicher1986transient,ivchenko1988magneto,pershoguba2022direct}.
More details of the partial agreement with some of these references but also the corrections of imprecisions and incorrect statements in some of them can be found in Appendix~\ref{appendixE}.

\begin{acknowledgments}
We would like to thank Yugo Onishi, Naoto Nagaosa, Liang Fu, Victor Yakovenko, Sergey Ganichev, Binghai Yan, and Fernando de Juan for stimulating discussions and correspondence.
We are particularly thankful to Mikhail Glazov and Leonid Golub for patiently 
and openly discussing with us their views on Ref.~\cite{belinicher1986transient} and also some of the subtle aspects of the physics of in-gap rectification, from which we benefited and gained key insights for composing Appendix~\ref{appendixE}.
\end{acknowledgments}

\bibliography{ingap-rigorous}

\clearpage

\appendix

\renewcommand{\theequation}{\thesection-\arabic{equation}}
\renewcommand{\thefigure}{\thesection-\arabic{figure}}
\renewcommand{\thetable}{\thesection-\Roman{table}}


\onecolumngrid

\section{Linear conductivity in the DC limit}
\label{appendixA}
In this appendix we show additional details of the linear conductivity in the DC limit discussed in the main text. 
In the DC limit $\omega \to 0$ the linear conductivity [see Eq.~(\ref{sigma1full}) in the main text] becomes: 
\begin{align}
\lim_{\omega \to 0}
\sigma_\Gamma^{\alpha\beta} (\omega)
= \frac{1}{2} 
\int_{\bf k} 
& \partial_\alpha\partial_\beta \bar{\epsilon}_{\bf k}
\Big[
\frac{f_\Gamma(\bar{\epsilon}_{\bf k})}{\Gamma}
-\frac{\partial g_\Gamma(\bar{\epsilon}_{\bf k})}{\partial\bar{\epsilon}_{\bf k}}
\Big]
\nonumber\\
= \frac{1}{2}\int_{\bf k} 
& \partial_\alpha\partial_\beta \bar{\epsilon}_{\bf k} 
\Big[
\frac{f_0(\bar{\epsilon}_{\bf k})}{\Gamma} 
+ 
\frac{\Gamma}{2}
\frac{\partial^3 f_0(\bar{\epsilon}_{\bf k})}{\partial \bar{\epsilon}_{\bf k}^3}
+
\frac{\Gamma^2}{3}
\frac{\partial^3 g_0(\bar{\epsilon}_{\bf k})}{\partial \bar{\epsilon}_{\bf k}^3}
+
O(\Gamma^3)
\Big] ,
\label{dc-linear}
\end{align}
in which
\begin{align}
g_\Gamma (\epsilon) = \frac{1}{2i} \Big[ f_+ (\epsilon) - f_- (\epsilon) \Big] ,
\quad
g_0 (\epsilon) \equiv \lim_{\Gamma \to 0} g_\Gamma(\epsilon) ,
\label{g_Gamma_2}
\end{align}
where $g_\Gamma (\epsilon)$ is the imaginary part of $f_+ (\epsilon)$ defined in Eq.~(\ref{fpm}) in the main text,
and we used the Cauchy–Riemann equations satisfied by $f_\Gamma(\epsilon)$ and $g_\Gamma(\epsilon)$
\begin{align}
& \frac{\partial f_\Gamma(\epsilon)}{\partial \Gamma}= \frac{\partial g_\Gamma(\epsilon)}{\partial \epsilon},
\quad
\frac{\partial f_\Gamma(\epsilon)}{\partial \epsilon}= - \frac{\partial g_\Gamma(\epsilon)}{\partial \Gamma} ,
\end{align}
and the resulting relation
\begin{align}
f_\Gamma(\epsilon)=f_0(\epsilon)+\Gamma\frac{\partial g_0 (\epsilon)}{\partial\epsilon} + O (\Gamma^2),
\label{f_gamma_g_0}
\end{align}
to obtaining the second equation of Eq.~(\ref{dc-linear}).
Therefore the clean limit of the DC conductivity resembles the prediction of the classic Drude theory for $\tau \equiv 1/(2\Gamma)$:
\begin{align}
\lim_{\Gamma \to 0}
\lim_{\omega \to 0}
\sigma_\Gamma^{\alpha\beta} (\omega)
=\frac{1}{2\Gamma}
\int_{\bf k} f_0(\bar{\epsilon}_{\bf k})\partial_\alpha\partial_\beta \bar{\epsilon}_{\bf k} ,
\end{align} 
and linear conductivity has a Drude peak in the DC limit when the chemical potential of the bath is within the bandwidth of the system $\mu\in[0,\Delta]$. 
The system can still have a finite linear DC conductivity even if the band is nominally fully empty or occupied at finite $\Gamma$, namely,
\begin{align}
& \lim_{\omega \to 0}
\sigma_\Gamma^{\alpha\beta} (\omega)
= 
\frac{1}{2}\int_{\bf k} \partial_\alpha\partial_\beta \bar{\epsilon}_{\bf k} 
\Big[
\frac{\Gamma^2}{3}
\frac{\partial^3 g_0(\bar{\epsilon}_{\bf k})}{\partial \bar{\epsilon}_{\bf k}^3}
+
O(\Gamma^3)
\Big]
\propto \Gamma^2 + O (\Gamma^3)  ,
\qquad
\big( T_0 \to 0 ,\
\mu \notin [0,\Delta] \big).
\end{align}
This conductance vanishes when $\Gamma \to 0$.

\section{Rectification conductivity in the DC limit}
\label{appendixB}
In this appendix we show more details of the rectification conductivity in the DC limit discussed in the main text. 
In the DC limit, the rectification conductivity [see Eq.~(\ref{sigma2full}) in the main text] is: 
\begin{align}
\lim_{\omega \to 0}
\sigma_{\Gamma}^{\gamma\alpha\beta}(\omega,-\omega) 
& =\frac{1}{4}
\int_{\bf k}\partial_\alpha\partial_\beta\partial_\gamma \bar{\epsilon}_{\bf k}
\Big[ \frac{f_\Gamma(\bar{\epsilon}_{\bf k})}{\Gamma^2} -\frac{1}{\Gamma}\frac{\partial g_\Gamma(\bar{\epsilon}_{\bf k})}{\partial\bar{\epsilon}_{\bf k}}-\frac{1}{3}\frac{\partial^2 f_\Gamma(\bar{\epsilon}_{\bf k})}{\partial\bar{\epsilon}_{\bf k}^2}\Big]
\nonumber\\
& = \frac{1}{4}\int_{\bf k} \partial_\alpha\partial_\beta \partial_\gamma \bar{\epsilon}_{\bf k}
\Big[
\frac{f_0(\bar{\epsilon}_{\bf k})}{\Gamma^2}
+
\frac{1}{6}\frac{\partial^2 f_0(\bar{\epsilon}_{\bf k})}{\partial \bar{\epsilon}_{\bf k}^2}
+ 
\frac{\Gamma^2}{24} \frac{\partial^4 f_0(\bar{\epsilon}_{\bf k})}{\partial \bar{\epsilon}_{\bf k}^4}
+ 
\frac{\Gamma^3}{45}
\frac{\partial^5 g_0(\bar{\epsilon}_{\bf k})}{\partial \bar{\epsilon}_{\bf k}^5}
+
O(\Gamma^4)
\Big],
\label{DC-rect-supp}
\end{align}
where we again used Eq.~(\ref{f_gamma_g_0}) in arriving at the second equation.
In the clean limit $\Gamma \to 0$, this coincides with the Jerk conductivity predicted within the relaxation time approximation, but here we also present the
sub-leading in $\Gamma$ correction:
\begin{align}
\lim_{\Gamma \to 0}
\lim_{\omega \to 0}
& \sigma_{\Gamma}^{\gamma\alpha\beta} (\omega,-\omega) 
=
\frac{1}{4}\int_{\bf k} 
\partial_\alpha\partial_\beta\partial_\gamma \bar{\epsilon}_{\bf k}
\Big[
\frac{f_0(\bar{\epsilon}_{\bf k})}{\Gamma^2}
+\frac{1}{6}\frac{\partial^2 f_0(\bar{\epsilon}_{\bf k})}{\partial \bar{\epsilon}_{\bf k}^2}
\Big] .
\label{clean-DC-rect}
\end{align}
Therefore, similarly to the linear conductivity, second order rectification conductivity has a Jerk peak at DC limit when the chemical potential is within the bandwidth of the system $\mu\in[0,\Delta]$.
When the band is nominally fully empty or occupied, for the rectification conductivity we now have
\begin{align}
\lim_{\omega \to 0}
&\sigma_\Gamma^{\gamma\alpha\beta} (\omega,-\omega)
=\frac{1}{4}\int_{\bf k} \partial_\alpha\partial_\beta\partial_\gamma \bar{\epsilon}_{\bf k}
\Big[
\frac{\Gamma^3}{45}
\frac{\partial^5 g_0(\bar{\epsilon}_{\bf k})}{\partial \bar{\epsilon}_{\bf k}^5}
+
O(\Gamma^4)
\Big]
\propto 
\Gamma^3 + O (\Gamma^4) ,
\qquad
\big( T_0 \to 0 ,\
\mu \notin [0,\Delta] \big).
\end{align}
This finite DC rectification conductivity again vanishes in the clean limit $\Gamma \to 0$.

\section{Second harmonic generation}
\label{appendixC}
In this appendix we show the second harmonic conductivity mentioned in the main text. 
The second harmonic conductivity is the one that controls the response oscillating at the double frequency of the drive ($s = 2$), we define it as:
\begin{align}
j_\gamma^{(2)} = \sigma_{\Gamma}^{\gamma\alpha\beta}(\omega,\omega)E^\alpha_\omega E^\beta_\omega + O(|E_\omega|^3),
\end{align}
and it is given by the following expression:
\begin{align}
\sigma_{\Gamma}^{\gamma\alpha\beta}(\omega,\omega) 
& =
-\frac{1}{2\omega^2}
\int_{\bf k}
f_\Gamma \partial_\alpha\partial_\beta\partial_\gamma \bar{\epsilon}_{\bf k}
-
\frac{1}{\omega^3}\frac{\Gamma}{2\Gamma - i\omega}\int_{\bf k} (\partial_\alpha\bar{\epsilon}_{\bf k})(\partial_\beta\partial_\gamma \bar{\epsilon}_{\bf k}){\cal L}_1(\bar{\epsilon}_{\bf k},\omega)
\nonumber\\
& \quad - \frac{1}{2\omega^4}\frac{\Gamma}{2\Gamma-2i\omega}\int_{\bf k} (\partial_\gamma\bar{\epsilon}_{\bf k})
\Big[
(\partial_\alpha \bar{\epsilon}_{\bf k})(\partial_\beta \bar{\epsilon}_{\bf k}){\cal L}_2(\bar{\epsilon}_{\bf k},\omega)
+\frac{\omega}{2}\partial_\alpha\partial_\beta \bar{\epsilon}_{\bf k}{\cal L}_1(\bar{\epsilon}_{\bf k},2\omega)\Big],
\end{align}
where
\begin{align}
{\cal L}_2(\bar{\epsilon}_{\bf k},\omega)=f_+(\bar{\epsilon}_{\bf k}-2\omega)-2f_+(\bar{\epsilon}_{\bf k}-\omega)+f_+(\bar{\epsilon}_{\bf k})+f_-(\bar{\epsilon}_{\bf k})-2f_-(\bar{\epsilon}_{\bf k}+\omega)+f_-(\bar{\epsilon}_{\bf k}+2\omega) .
\end{align}

The low frequency limit of second harmonic conductivity coincides with the low frequency limit of rectification conductivity from Eq.~(\ref{DC-rect}) in the main text:
\begin{align}
 \lim_{\omega \to 0}
\sigma_{\Gamma}^{\gamma\alpha\beta}&(\omega,\omega) =\frac{1}{4}
\int_{\bf k} \partial_\alpha\partial_\beta\partial_\gamma \bar{\epsilon}_{\bf k}
\Big[ \frac{f_\Gamma(\bar{\epsilon}_n)}{\Gamma^2} -\frac{1}{\Gamma}\frac{\partial g_\Gamma(\bar{\epsilon}_n)}{\partial\bar{\epsilon}_n}-\frac{1}{3}\frac{\partial^2 f_\Gamma(\bar{\epsilon}_n)}{\partial\bar{\epsilon}_n^2}\Big].
\end{align}
Interestingly, at large frequencies $\omega\gg\Delta$ the real part of the second harmonic conductivity decays as $1/\omega^2$ in contrast to $1/\omega^4$ power decay of the rectification conductivity.

\section{Relation to the Boltzmann theory}
\label{appendixD}

In this appendix we discuss the relation between our result and that from a simpler Boltzmann/relaxation-time approach.
We begin by writing a Boltzmann equation for a single band system in the relaxation time approximation:
\begin{align}
\partial_t f ({\bf k}, t) + {\bf E}(t) \cdot \nabla_{\bf k} f ({\bf k}, t) 
= - [ f ({\bf k}, t) - f_0 ( \epsilon_{\bf k}) ] / \tau ,
\label{boltzmann-gauge-0}
\end{align}
where ${\bf E}(t)= {\bf E}_\omega e^{-i \omega t} + \text{c. c.}$ is a monochromatic electric field.

The above equations are written in a different gauge with respect to the main text: here $\bf k$ is viewed as a gauge invariant mechanical crystal momentum, which corresponds to ${\bf k}-{\bf A}(t)$ in the main text.
In order to obtain expressions for occupation functions in the same gauge as in the main text, we convert to a gauge in which we keep track of the occupation of canonical crystal momenta, using the following relation:
\begin{align}
p ({\bf k},t) \equiv f ({\bf k}-{\bf A}(t),t) .
\end{align}
The occupation function $p ({\bf k},t)$ satisfies the following equation
\begin{align}
\partial_t p ({\bf k}, t) & = \partial_t f ({\bf k} - {\bf A}(t), t) - \partial_t  {\bf A}(t) \cdot \nabla_{\bf k} f ({\bf k} - {\bf A}(t), t)
\nonumber\\
& = \partial_t f ({\bf k} - {\bf A}(t), t) + {\bf E}(t) \cdot \nabla_{\bf k} f ({\bf k} - {\bf A}(t), t)
\nonumber\\
& = - [ f ({\bf k} - {\bf A}(t) , t) - f_0 ( \epsilon_{ {\bf k} - {\bf A}(t) } ) ] / \tau,
\end{align}
where we used Eq.~(\ref{boltzmann-gauge-0}) in obtaining the last equation. Therefore we see that the distribution function $p ({\bf k}, t) $ satisfies an equation without explicit electric field
derivative term:
\begin{align}
\partial_t p ({\bf k}, t) = - [ p ({\bf k},t) - f_0 ( \epsilon_{ {\bf k} - {\bf A}(t) } ) ] /\tau .
\label{boltzmann-gauge-1}
\end{align} 
Using the fact that the late-time steady state distribution is periodic,
we perform Fourier series expansions for both $p ({\bf k},t) $ and $f_0 ( \epsilon_{ {\bf k} - {\bf A}(t) } )$:
\begin{align}
& p ({\bf k}, t) = \sum_{l=-\infty}^{+\infty} p^{(l)} ({\bf k}) \exp(-i l \omega t) ,
\quad
p^{(l)} ({\bf k}) = \int_0^T \frac{{\text d} t}{T} 
p ({\bf k}, t) \exp( i l \omega t) ;
\nonumber\\
& f_0 ( \epsilon_{ {\bf k} - {\bf A}(t) } ) = \sum_{l=-\infty}^{+\infty} f_0^{(l)} ({\bf k}) \exp(-i l \omega t) ,
\quad
f_0^{(l)} ({\bf k}) = \int_0^T \frac{{\text d} t}{T} 
f_0 ( \epsilon_{ {\bf k} - {\bf A}(t) } ) \exp(i l \omega t) .
\end{align}
With the above expansions, Eq.~(\ref{boltzmann-gauge-1}) becomes
\begin{align}
- i l \omega p^{(l)} ({\bf k}) =  - p^{(l)} ({\bf k}) / \tau + f_0^{(l)} ({\bf k}) / \tau, 
\end{align}
and leads to
\begin{align}
p^{(l)} ({\bf k}) = \frac{1}{1 - i l \omega \tau} f_0^{(l)} ({\bf k}) .
\end{align}
The above solution in general requires an explicit calculation of the following mixed harmonics of the distribution:
\begin{align}
f_0^{(l)} ({\bf k}) = \int_0^T \frac{{\text d} t}{T} 
f_0 ( \epsilon_{\bf k}^{(0)} + \epsilon_{\bf k}^{(1)} e^{-i \omega t}+ \epsilon_{\bf k}^{(-1)} e^{ i \omega t} + \cdots ) \exp(i l \omega t) ,
\end{align}
where
\begin{align}
\epsilon_{ {\bf k} - {\bf A}(t) } =  \sum_{l=-\infty}^{+\infty}  \epsilon_{\bf k}^{(l)}  \exp(-i l \omega t) ,
\quad
\epsilon_{\bf k}^{(l)} = \int_0^T \frac{{\text d} t}{T} 
\epsilon_{ {\bf k} - {\bf A}(t) } \exp(i l \omega t) .
\end{align}
Let us consider however the clean limit $\tau \to + \infty$. Notice that $f_0^{(l)} ({\bf k})$ is independent of $\tau$, therefore for $l \neq 0$ components we have
\begin{align}
\lim_{\tau \to + \infty} p^{(l \neq 0)} ({\bf k}) = \lim_{\tau \to + \infty}  \frac{1}{1 - i l \omega \tau} f_0^{(l)} ({\bf k}) = 0.
\end{align}
However the $l=0$ component, or time averaged component, which is independent of $\tau$ and therefore remains finite as $\tau \to + \infty$, is given by:
\begin{align}
p^{(0)} ({\bf k}) =  f_0^{(0)} ({\bf k}) = \int_0^T \frac{{\text d} t}{T} 
f_0 ( \epsilon_{\bf k}^{(0)} + \epsilon_{\bf k}^{(1)} e^{-i \omega t}+ \epsilon_{\bf k}^{(-1)} e^{ i \omega t} + \cdots ) .
\label{p0-boltzmann}
\end{align}
Therefore, similarly to Eq.~(\ref{p_k_Gamma0}) obtained from the full formalism with the bath, the distribution from the Boltzmann theory becomes time independent in the canonical crystal momentum, but not in the mechanical physical momentum, in the analogous ideal limit of $\tau \to + \infty$.
Notice, however, that the above result has to be viewed as a limit of $\tau \to + \infty$, and not as a situation in which there is no relaxation.
This is because in the strict absence of relaxation mechanisms there is no unique late-time steady state, namely by taking $1/\tau = 0$ and neglecting altogether the relaxations in the right hand side of Eq.~(\ref{boltzmann-gauge-1}) any time-independent distribution of the canonical momenta would be a solution.

If we expand up to the second order of electric fields Eq.~(\ref{p0-boltzmann}) we obtain:
\begin{align}
p^{(0)} ({\bf k})  & = \int_0^T \frac{{\text d} t}{T} 
\Big[
f_0 ( \bar{\epsilon}_{\bf k} )
+
\big( \epsilon_{\bf k}^{(1)} e^{-i \omega t}+ \epsilon_{\bf k}^{(-1)} e^{ i \omega t} + 
\epsilon_{\bf k}^{(2)} e^{- i 2 \omega t}+ \epsilon_{\bf k}^{(-2)} e^{ i 2 \omega t} \big)
f_0^\prime ( \bar{\epsilon}_{\bf k} )
\nonumber\\
& \hspace{5em} + \frac{1}{2} \big( \epsilon_{\bf k}^{(1)} e^{-i \omega t}+ \epsilon_{\bf k}^{(-1)} e^{ i \omega t} \big)^2 f_0^{\prime \prime} ( \bar{\epsilon}_{\bf k} ) + O (|E_\omega|^3)
\Big]
\nonumber\\
& = f_0 ( \bar{\epsilon}_{\bf k} ) + 
|\epsilon_{\bf k}^{(1)}|^2 f_0^{\prime \prime} ( \bar{\epsilon}_{\bf k} ) + O (|E_\omega|^3) .
\end{align}
Interestingly the above distribution function coincides with the asymptotic behavior of the staircase distribution function discussed in the main text [see e.g.,
Eq.~(\ref{ideal-occupation-2nd})] in limit of $\Delta \gg \omega \gg \Gamma \to 0$:
\begin{align}
\lim_{\omega \to 0}
\lim_{\Gamma \to 0} p_{\bf k} & = 
\lim_{\omega \to 0} 
\Big[
\big( 1 - \frac{2\big|\epsilon_{\bf k}^{(1)}\big|^2}{\omega^2} \big) f_0 ( \bar{\epsilon}_{\bf k})
+ \frac{\big|\epsilon_{\bf k}^{(1)}\big|^2}{\omega^2} f_0 ( \bar{\epsilon}_{\bf k} - \omega) 
+ \frac{\big|\epsilon_{\bf k}^{(-1)}\big|^2}{\omega^2} f_0 ( \bar{\epsilon}_{\bf k} + \omega) 
\Big]
= f_0 (\bar{\epsilon}_{\bf k}) + |\epsilon_{\bf k}^{(1)}|^2
f_0^{\prime \prime} ( \bar{\epsilon}_{\bf k} ) .
\label{full-boltz-consistent}
\end{align}
Therefore the expectation value of all equal time observables, such as the electric current, coincide with those of the more microscopic Floquet-bath theory of the main text, at least to second order in electric fields. In particular one obtains the same rectification conductivity in the above limit as that in Eq.~(\ref{dc-clean-rec}) of the main text, that we refer to as Jerk effect.

\section{Comments and connections to other works in the literature}
\label{appendixE}

There has been a long-standing debate in the literature about the possibility of in-gap rectification which has been clouded by previous imprecise and incorrect statements.
In this section we will try to clarify some of this. We begin by defining precisely what do we mean by in-gap rectification.
The optical gap is defined as the region in the frequency domain in which the the hermitian symmetric part of the conductivity tensor vanishes in the limit of low temperatures and small scattering rates (see Ref.~\cite{shi2022berry} for a review).
We then say that a system has in-gap rectification if any of the elements of the  rectification conductivity tensor that lead to finite DC currents generated by a monochromatic AC electric field with a frequency within the optical gap remain non-zero in that same limit. More specifically:
\begin{align}
& \text{Definition of  ``optical gap'' :}
\qquad
\lim_{T_0 \to 0} \lim_{\Gamma \to 0} \big( \sigma^{\alpha \beta} (\omega)+ [\sigma^{\beta \alpha} (\omega)]^* \big) \to 0, 
\quad
\text{when $\omega \in$ optical gap}.
\\
& \text{Definition of  ``in-gap rectification'' :}
\qquad
\lim_{T_0 \to 0} \lim_{\Gamma \to 0}  \sigma^{\gamma \alpha \beta} (\omega,-\omega) \neq 0, 
\quad 
\text{when $\omega \in$ optical gap}.
\end{align}
Therefore our current manuscript and our previous work in Ref.~\cite{shi2022berry}, demonstrate rigorously that in-gap rectification in the above sense is indeed possible. 

Nevertheless, some confusion in the literature appears to have originated from different interpretations of the work of Belinicher, Ivchenko, and Pikus (BIP) in Ref.~\cite{belinicher1986transient}. 
That paper contained statements such as ``The conclusion that a steady-state photocurrent may appear on illumination in the transparency range of a crystal, reached in earlier publications, is shown to be in error''. 
This statement could be read as implying the impossibility of in-gap rectification in the sense we defined above.
In fact, this reading of the BIP paper appears to have been made in several references claiming that in-gap rectification in the above sense is impossible~\cite{onishi2022effects,de2020difference,pershoguba2022direct}.
Even us in our recent work of Ref.~\cite{shi2022berry}, read the BIP paper as trying to prove that in-gap rectification is impossible in the above sense.

However, part of the issue with reading the aforementioned BIP paper, is that it left several crucial gaps in its discussion and its derivations that can make it hard to know in a precise way what exactly BIP implied at various places and the precise framework that BIP used for reaching such conclusions.
For example, a crucial point that can lead to a different readings of the BIP paper relates to the definition of the term ``$g_n$'' that appears in the right hand side of their Eq.~(8) in Ref.~\cite{belinicher1986transient}, which is a central equation from which various conclusions are derived.
Unfortunately BIP never spelled out an explicit form for this term, but simply wrote that ``$g_n$ is the generation function, i.e., the rate of change of the distribution function due to optical transitions.''.
This leaves open to interpretation what exactly they had in mind for ``optical transitions''.
For example, one could read this by interpreting ``$g_n$'' as associated only with inter-band optical transitions, and in this case, one would be lead to read the BIP paper as trying to imply that in-gap rectification in the above sense is impossible.

There is however an alternative way to interpret ``$g_n$'' and the notion of ``optical transitions'' in Ref.~\cite{belinicher1986transient} as a more general notion of irreversible ``transitions'' that can take place even within what would nominally be the optical gap defined in the above sense.
This more nuanced way of interpreting the BIP paper has indeed been recently emphasized by Glazov and Golub in Ref.~\cite{golub2022raman}. For example Golub and Glazov write in Ref.~\cite{golub2022raman} that ``... even for transparent media, real electronic transitions should occur to enable the photocurrent.'' and that ``We reiterate that in the absence of any real electronic transitions DC current is forbidden. It is obvious from general reasons: If a DC current is generated then this current results in a Joule heat in the sample or in the external circuit connected to the sample. It is forbidden by the energy conservation law in the absence of real transitions.''. 
What Golub and Glazov are trying to explain there is in line with our recent thermodynamic analysis in Ref.~\cite{shi2022berry}, where we emphasized that in order to guarantee the positivity of entropy production, specially when the system is connected to an external circuit, it is always important to view the scattering rate $\Gamma$ as possibly arbitrarily small but not strictly zero.
This requirement means that physically it is important to have always a non-zero absorption within the nominal optical gap of the material. 
In fact Golub, Ivchenko himself and Spivak, have also emphasized a related aspect of this in Ref.~\cite{golub2020semiclassical} where they demonstrated that the CPGE effect associated with the Berry dipole term remains finite within the optical in the limit of $\Gamma \to 0$, but also coexists the other contributions that originate from impurity scattering mechanisms that scale in the same way with frequency and remain finite inside of the gap in the limit of $\Gamma \to 0$.
One way to state this state of affairs, that has been emphasized by Golub and Glazov to us in private communications, is that while the real transitions associated with scattering lead to a vanishingly small linear dissipative conductivity in the limit of $\Gamma \to 0$, there are cancellations of the scattering rate that lead to finite rectification conductivity in this limit but the ``real electronic transitions'' are still taking place. 
These ``real electronic transitions'' are therefore the more general notion of ``optical transitions'' that can contribute to the term ``$g_n$'' in the BIP reference.
Therefore, within this point of view, one can say that the BIP should not be read as implying that in-gap rectification is impossible in the sense we defined above.
We are in agreement with the physics of this point of view broadly speaking.

There is however another crucial aspect of the BIP work in Ref.~\cite{belinicher1986transient} with which we still find ourselves in disagreement and that we believe our current paper provides good evidence to be incorrect in general. 
BIP stated that ``... in the case of continuous illumination the steady-state distribution function is $f_0( \bar{\epsilon}_{\bf k})$ irrespective of how weak is the interaction of electrons with phonons.''
In this statement $f_0$ is the ``equilibrium distribution function'' (the Fermi-Dirac occupation function) and $\bar{\epsilon}_{\bf k}$ is the Floquet energy of the band. These statement has been echoed in several subsequent works~\cite{onishi2022effects,golub2022raman,ivchenko1988magneto,pershoguba2022direct}.
However, our current work demonstrates that in the limit of $\Gamma \to 0$ the distribution function is sharply different from the naive Fermi-Dirac occupation, but becomes instead the non-trivial Fermi-Dirac staircase discussed in the main text, even to the leading order $|E_\omega|^2$ in the driving monochromatic field.
Crucially the resulting occupation function cannot be expressed as a function of the Floquet energy alone [see Fig.~\ref{fig-ladder}, Eq.~(\ref{p_k_Gamma0}), and Eq.~(\ref{ideal-occupation-2nd}) of the main text].
Notice that in order to have a unique and well defined steady state at late times, we must necessarily view the relaxation rate $\Gamma$ as being arbitrarily small but not strictly zero.
Therefore, the notion of the ideal occupation in the steady state has to be necessarily interpreted as the limit of $\Gamma \to 0$ of the occupation of systems with a finite $\Gamma$. 
This is because systems with strictly zero relaxation rate ($\Gamma = 0$) do not have a way to erase the memory of their initial conditions and therefore their steady state in the presence of the monochromatic light is not uniquely defined. 

We have demonstrated rigorously that at least for an ideal fermionic bath the occupation of states in the limit of $\Gamma \to 0$ is not $f_0(\bar{\epsilon}_{\bf k})$ as Refs.~\cite{onishi2022effects,golub2022raman,belinicher1986transient,ivchenko1988magneto,pershoguba2022direct} presumed.
We would like to emphasize that while the fermionic bath might appear to be a somewhat artificial approximation to the true mechanisms of relaxation for certain realistic physical situations, it behaves as an ideal thermal bath in the limit $\Gamma \to 0$.
In particular, the particle number becomes effectively conserved in such limit since the self-consistent occupations at each momentum become a time independent function as we have shown. 
We have in particular demonstrated that in equilibrium this bath leads to the expected Fermi-Dirac occupation of the system. 
More generally speaking, in equilibrium one expects a universality of all intensive thermodynamic physical properties of the system of interest for a large class of baths regardless of their details, which essentially defines the class of ``ideal thermodynamic baths''.
However, how this universality carries over to non-equilibrium settings is still unclear to us.
Therefore whether other baths or other relaxation mechanisms such as coupling to phonons, impurities or self-thermalization via electron-electron interactions lead to a similar stair-case occupation to the one we have found in the limit of vanishing relaxation rates $\Gamma \to 0$, remains an interesting open problem.
We note however that none of the aforementioned Refs.~\cite{onishi2022effects,golub2022raman,belinicher1986transient,ivchenko1988magneto,pershoguba2022direct} has provided a rigorous and controlled derivation of the self-consistent occupation of Floquet bands based on any microscopically explicit mechanism of relaxation, like the one we have provided. 
Therefore we do not see any rigorous substance to their claim that the occupation is $f_0(\bar{\epsilon}_{\bf k})$ even for other microscopic relaxation mechanisms such as phonons.
Moreover, it has become abundantly clear in the study of thermalization of Floquet systems in recent years that the self-consistent occupation of Floquet bands coupled to baths that are also bosonic differs clearly from the naive Fermi-Dirac distribution of the Floquet bands $f_0 (\bar{\epsilon}_{\bf k})$~\cite{bilitewski2015scattering,shirai2015condition,seetharam2015controlled,genske2015floquet,shirai2016effective,esin2018quantized}.

\end{document}